\documentclass[12pt]{article}

\pdfoutput=1

\usepackage{graphicx}
\usepackage{epsfig}
\usepackage{amsfonts}
\usepackage{amssymb}
\usepackage{amsmath}
\usepackage{dsfont}
\usepackage{bbm}
\usepackage{multirow}
\usepackage[vcentermath]{youngtab}
\usepackage{latexsym}
\usepackage{verbatim}
\usepackage{mcite}
\usepackage{cite}
\usepackage{units}

\def\beq{\begin{equation}}
\def\eeq{\end{equation}}
\def\beqa{\begin{eqnarray}}
\def\eeqa{\end{eqnarray}}

\def\a{{\alpha}}
\def\b{{\beta}}

\def\bfone{\relax{\rm 1\kern-.35em 1}}

\newcommand{\cN}{{\cal N}}

\newcommand{\be}{\begin{equation}}
\newcommand{\ee}{\end{equation}}
\newcommand{\ben}{\begin{displaymath}}
\newcommand{\een}{\end{displaymath}}
\newcommand{\bea}{\begin{eqnarray}}
\newcommand{\eea}{\end{eqnarray}}

\newcommand{\bean}{\begin{eqnarray*}}
\newcommand{\eean}{\end{eqnarray*}}

\begin{document}
\pagestyle{plain}


\makeatletter
\@addtoreset{equation}{section}
\makeatother
\renewcommand{\thesection}{\arabic{section}}
\renewcommand{\theequation}{\thesection.\arabic{equation}}
\renewcommand{\thefootnote}{\arabic{footnote}}


\setcounter{page}{1}
\setcounter{footnote}{0}


\begin{titlepage}
\begin{flushright}
\small ~~
\end{flushright}

\bigskip

\begin{center}

\vskip 0cm

{\LARGE \bf Charting the landscape\\[5mm] of $\,\mathcal{N}=4\,$ flux compactifications} \\[6mm]
\vskip 0.5cm

{\bf Giuseppe Dibitetto,\, Adolfo Guarino \,and\, Diederik Roest}\\

\vskip 25pt

{\em Centre for Theoretical Physics,\\
University of Groningen, \\
Nijenborgh 4, 9747 AG Groningen, The Netherlands\\
{\small {\tt \{g.dibitetto , j.a.guarino , d.roest\}@rug.nl}}} \\

\vskip 0.8cm

\end{center}

\vskip 1cm

\begin{center}

{\bf ABSTRACT}\\[3ex]

\begin{minipage}{13cm}
\small

We analyse the vacuum structure of isotropic $\mathbb{Z}_2 \times \mathbb{Z}_2$ flux compactifications, allowing for a single set of sources. Combining algebraic geometry with supergravity techniques, we are able to classify all vacua for both type IIA and IIB backgrounds with arbitrary gauge and geometric fluxes. Surprisingly, geometric IIA compactifications lead to a unique theory with four different vacua. In this case we also perform the general analysis allowing for sources compatible with minimal supersymmetry. Moreover, some relevant examples of type IIB non-geometric compactifications are studied. 
The computation of the full $\cN = 4$ mass spectrum reveals the presence of a number of
non-supersymmetric and nevertheless stable AdS$_4$ vacua. In addition we find a novel dS$_4$ solution based on a non-semisimple gauging.
\end{minipage}

\end{center}

\vfill

\end{titlepage}


\tableofcontents

\section{Introduction}

Since the turn of the millenium, a lot of progress has been made in
the context of flux compactifications of string theory in order to
obtain four-dimensional effective descriptions with a number of
desired features. In particular, from a phenomenological point of
view, one is interested in a vacuum with small but positive
cosmological constant and spontaneously broken supersymmetry. This
implies the necessity of finding de Sitter (dS) solutions from string
theory compactifications. In addition to modelling dark energy, these are relevant for embedding
descriptions of inflation in string theory. Moreover, Anti-de Sitter (AdS) solutions are employed in holographic
applications in order to study physical systems which have a
conformal symmetry realised in the UV.

Many string theory constructions related to flux backgrounds compatible
with minimal supersymmetry have been studied so far. In 
particular, the mechanism of inducing an effective
superpotential from fluxes \cite{Giddings:2001yu} has been extensively studied in the
literature for those compactifications giving rise to a
so-called $STU$-model as low energy description \cite{Kachru:2002he,Derendinger:2004jn,DeWolfe:2004ns,Camara:2005dc,Villadoro:2005cu,Derendinger:2005ph,DeWolfe:2005uu,Aldazabal:2006up,Aldazabal:2007sn}. However, recent
progress in understanding the link between half-maximally
supersymmetric string backgrounds and gaugings of $\mathcal{N}=4$
supergravity \cite{Aldazabal:2008zza,Dall'Agata:2009gv,Dibitetto:2010rg}, seems to give a
powerful tool for addressing the same issue in the context of
$\mathcal{N}=4$ compactifications. As we will discuss later, this allows one to address the stabilisation of all
moduli consistent with the isotropic $\mathbb{Z}_2 \times \mathbb{Z}_2$ orbifold compactification.

Another interesting opportunity offered by the study of
such flux compactifications and their relation to half-maximal
supergravity, is that of addressing the issue of stability without
supersymmetry in extended supergravity. More precisely, for a long
time it was believed that there are no stable vacua of maximal or half-maximal supergravity that spontaneously break all supersymmetry. Very recently
\cite{Fischbacher:2010ec}, however, an example of an AdS
critical point which is both non-supersymmetric and stable 
has been found in maximal supergravity. This adds further motivation to look for new such
extrema in the half-maximal case as well. Furthermore, the
possible existence of stable de Sitter vacua in this context still
remains an open discussion point \cite{Borghese:2010ei}.

In maximal supergravity with SO$(8)$ gauge group, the main approach to classify critical points has been to consider a particular truncation, restricting only to the degrees of freedom that are singlets with respect to a certain symmetry group, e.g.~an SU$(3)$ subgroup of SO$(8)$. The consistency of the truncation ensures the extremality of the non-singlet scalars that are truncated out. However, it by no means implies any restriction on the mass of these scalars, and hence in order to check e.g.~stability of a particular critical point, one should consider the full theory. A striking example is provided by a particular critical point of $\cN = 8$ supergravity that is invariant under SU$(4)^-$: even though all singlet scalars are stable, there are instabilities in the non-singlet sector \cite{Warner:1006}. This underlines the importance of considering the mass spectrum of the full theory. We will adopt a similar approach towards the classification of critical points of general $\cN = 4$ theories, by requiring the critical points to preserve at least an SO$(3)$ subgroup of the gauge group. This will allow us not only to classify the different critical points of a particular theory, but also all the theories that allow for moduli stabilisation in e.g.~geometric IIA compactifications.

With respect to the string theory interpretation of the theories at hand, progress in this direction has been (partially) motivated by the search for de Sitter solutions. Firstly, a \emph{no-go} result was proven which rules out the possibility of having de Sitter solutions in the presence of only gauge fluxes
\cite{Hertzberg:2007wc}. Further generalisations have investigated the possibility to circumvent this no-go theorem
by including geometric fluxes, see e.g.~\cite{Silverstein:2007ac,Flauger:2008ad,Haque:2008jz,Caviezel:2008tf,Danielsson:2009ff,deCarlos:2009fq,Caviezel:2009tu}. 
However, the difficulties in finding de Sitter solutions in an $\cN = 4$ set-up with only gauge fluxes and geometric fluxes \cite{Dibitetto:2010rg}, make it necessary to go beyond those
ingredients. A first extension has been carried out by introducing
the so-called T-folds in doubled geometry \cite{Hull:2006va,
Hull:2007jy}; this is a T-duality-covariant construction obtained by
supplementing the internal space with extra coordinates conjugate to
winding number. A second extension goes towards the introduction
of non-geometric fluxes. These were introduced
as dual counterparts of geometric and gauge fluxes based on mirror
symmetry \cite{Shelton:2005cf, Grana:2006hr}, thus allowing for the generalisation of
duality symmetries in the presence of fluxes. This construction
turns out to be natural in the context of type IIB string theory. However, the relation between these two
generalisations of a flux background is not completely immediate and
turns out to depend on the duality frame. In the present paper we will mainly focus on
gauge and geometric fluxes, and only lightly touch upon some non-geometric fluxes.

The paper is organised as follows. In section \ref{sec:sugra}, we
first review the embedding tensor formulation of half-maximal
supergravity theories and discuss the structure of the underlining
gauging; secondly we construct an SO($3$) truncation thereof and
interpret it in the $\mathcal{N}=1$ superpotential language which
allows us to spell out a complete dictionary between fluxes and
embedding tensor components. In section \ref{sec:crit}, we present the tools used
in order to analyse critical points and discuss their features. In
section \ref{sec:geom_IIA}, we present the complete set of vacua of
geometric type IIA $\mathcal{N}=4$ compactification. In section
\ref{sec:nongeom_IIB}, we give the complete set of vacua of type IIB
compactifications with only gauge fluxes and some relevant solutions
of non-geometric type IIB compactifications. Finally we present our conclusions in section 6. In appendix~\ref{App:N1_vacua}
we present the classification of vacua in the case of geometric type IIA $\cN =1$ compactifications.

\section{$\,\cN=4\,$ supergravities from flux compactifications}
\label{sec:sugra}

In this section we present a brief introduction to half-maximal supergravity theories in four dimensions. We will focus on those arising as consistent SO($3$) truncations of the general theory and will show that they admit a string theory realisation in terms of flux compactifications in the presence of generalised background fluxes. 

\subsection{General review of $\,\cN=4\,$ gauged supergravities}

We mostly follow the notation and conventions of ref.~\cite{Schon:2006kz} to work out the $\cN=4$ supergravity theory invariant under the action of the $G$ = SL($2$) $\times$ SO($6,6$) duality group in four dimensions.

\subsubsection*{Gauge vectors and gauge algebra}

The theory contains vector fields $A_{\mu}$ in four dimensions which transform in the fundamental representation of $G$ = SL($2$) $\times$ SO($6,6$),
\beq
\begin{array}{cc}
A_{\mu}= V_{\mu}^{\a M} \, T_{\a M} \ ,
\end{array}
\eeq
where $\,\a=(+,-)\,$ is a fundamental SL($2$) index and
$\,M=1,...,12\,$ is the SO($6,6$) fundamental index.

In the \textit{ungauged} theory, only a subgroup
$G_{0}=\textrm{U}(1)^{12} \subset \textrm{SO}(6,6)$ is realised and
the vector fields become abelian, i.e. $\left[  T_{\a M} , T_{\b N}
\right]=0$. However, this \textit{ungauged} theory can be deformed
away from the abelian structure without breaking the $\,\cN=4\,$
supersymmetry so that a non-abelian subgroup $G_{0} \subset
\textrm{SO}(6,6)$ is realised \cite{Schon:2006kz}. Then, the most general form of the
gauge algebra in the \textit{gauged} theory becomes\footnote{In general this can be extended with deformation parameters $\xi_{\alpha M}$. We will not include these here as such  parameters are completely projected out in  
the SO$(3)$ truncation that we analyse in the present paper. \label{foot_xi}}
\beq
\label{SW_algebra}
\begin{array}{ccc}
\left[  T_{\a M} , T_{\b N} \right] &=&  {f_{\a\,MN}}^{P} \,\,T_{\b P} \ ,
\end{array}
\eeq
with $f_{\a\,MNP}={f_{\a\,MN}}^{Q} \, \eta_{QP} = f_{\a\,[MNP]}\,$
being the structure constants of $\,G_{0}\,$ and with
$\,\eta_{MN}\,$ the SO($6,6$) metric. This automatically implies
that only the $G_{0} \subset \textrm{SO}($6,6$)$ subgroups admitting
 $\eta_{MN}$ as a non-degenerate bi-invariant metric can be
realised as deformations of the ungauged theory. In other words, the
adjoint representation of $G_{0}$ has to be embeddable within the
fundamental representation of SO($6,6$). This embedding may not be
unique, thus resulting in non-equivalent realisations of the same $G_{0}$
subgroup. From now on, we will use light-cone coordinates, so that
an SO($6,6$) index is raised or lowered by using the SO($6,6$)
light-cone metric
\beq
\label{eta_lc}
\eta_{MN}=\eta^{MN}=\begin{pmatrix} 0 & \mathds{I}_{6}  \\ \mathds{I}_{6} & 0 \end{pmatrix} \ .
\eeq

Let us perform the splitting of the fundamental SO($6,6$) index $\,M
\equiv ({}_{m} \, , \, {}^{m})\equiv (m,\bar{m})\,$ with
$\,m=1,...,6\,$ and $\,\bar{m}=\bar{1},...,\bar{6}\,$. Then, the
vectors split as $\,T_{\a M} \equiv (Z_{\a m} \, , \,
{X_{\a}}^{m})\,$ alike, and the algebra in (\ref{SW_algebra}) can be
rewritten as the set of brackets
\beq
\label{SW_algebra_ZX}
\begin{array}{cccccc}
\left[  Z_{\a m} , Z_{\b n} \right] &=&  {f_{\a\,mn}}^{p} \,\,Z_{\b p} &+& f_{\a\,mnp} \,\,{X_{\b}}^{p} & , \\[1mm]
\left[  Z_{\a m} , {X_{\b}}^{n} \right] &=&  {f_{\a\,m}}^{np} \,\,Z_{\b p} &+& f_{\a\, m\phantom{n}\,p}^{\phantom{\a}\phantom{m}n\phantom{p}} \,\,{X_{\b}}^{p} & , \\[1mm]
\left[ {X_{\a}}^{m}, Z_{\b n} \right] &=&  f^{\phantom{\a}m\phantom{n}p}_{\a\,\phantom{m}n\phantom{p}} \,\,Z_{\b p} &+& f_{\a\, \phantom{m}np}^{\phantom{\a} m\phantom{np}} \,\,{X_{\b}}^{p} & , \\[1mm]
\left[  {X_{\a}}^{m} , {X_{\b}}^{n} \right] &=&  {f_{\a\,}}^{mnp} \,\,Z_{\b p} &+& f_{\a\,\phantom{mn}p}^{\phantom{\a}mn\phantom{p}} \,\,{X_{\b}}^{p} & .\\[1mm]
\end{array}
\eeq
It is worth noticing that this is only apparently a
twenty-four-dimensional gauge algebra, but in fact the actual
gauging is twelve-dimensional after imposing the constraints 
 \begin{align}
  \epsilon^{\alpha \beta} \,f_{\alpha MNP} \, T_\beta{}^P  = 0 \,, \label{abstract-QC}
 \end{align}
which ensure the
anti-symmetry of the brackets in 
\eqref{SW_algebra}. This fact is related to the
observation in ref.~\cite{Dall'Agata:2007sr}, i.e. that the algebra
realised on the vectors can only be embedded in Sp($24$), whereas
the proper gauge algebra is that one realised on the curvatures,
which is obtained from the previous one after dividing out by the
abelian ideal consisting of all generators acting trivially on the
curvatures. To summarise, in order to identify the correct
gauging, one has to solve these constraints by expressing half of
the generators in terms the other ones and plug the solution into the brackets of
\eqref{SW_algebra_ZX}. 

\subsubsection*{Quadratic constraints and scalar potential}

The scalars of the theory span the coset geometry
\beq
\label{coset}
\frac{\textrm{SL}(2)}{\textrm{SO}(2)} \times \frac{\textrm{SO}(6,6)}{\textrm{SO}(6) \times \textrm{SO}(6)} \ .
\eeq
We will name $\,M_{\alpha\beta}\,$ the scalars parameterising the first factor and $\,M_{MN}\,$ those ones parameterising the second factor in (\ref{coset}). For the former we will use the following explicit parameterisation
\beq
\label{chi,phi}
M_{\alpha \beta} = e^{\phi} 
\left(
\begin{array}{cc} 
\chi^2 + e^{-2 \phi} & \chi \\ 
\chi & 1 
\end{array} 
\right) \hspace{5mm} , \hspace{5mm} \alpha = (+,-) \ ,
\eeq
where the SL$(2)$ indices are raised and lowered using $\,\epsilon_{\alpha \beta} = \epsilon^{\alpha \beta}\,$ with $\,\epsilon^{+-} = -\epsilon^{-+} = 1$. The matrix $\,M_{MN}$, can be determined by starting from a 'vielbein' denoted by $\,\mathcal{V}_{M}^{\phantom{M}\underline{A}}$, where $\,\underline{A}\,$ is an SO($6$) $\times$ SO($6$) index whereas $\,M\,$ is an SO($6,6$) one. This object is such that
\beq
\label{M=VV}
M=\mathcal{V} \, \mathcal{V}^{T} .
\eeq
Global SO$(6,6)$ transformations act on $\,\mathcal{V}\,$ from the left, whereas local $\textrm{SO}(6) \times \textrm{SO}(6)$ transformations act from the right. Even though $\,\mathcal{V}\,$ is not by itself invariant under local $\textrm{SO}(6) \times \textrm{SO}(6)$ transformations, the particular combinations constructed out of it which will appear in the scalar potential are. In particular, the matrix $\,M\,$ itself is invariant.

As for the embedding tensor components, they can be parameterised by $\,f_{\alpha [MNP]}\,$ and $\,\xi_{\alpha M}\,$, but, as discussed in footnote~\ref{foot_xi}, we will set $\,\xi_{\alpha M}=0\,$ in the following formula. The non-vanishing embedding tensor components $\,f_{\alpha MNP}\,$ have to satisfy the following quadratic constraints\footnote{The only further subtlety is that the second set of quadratic constraints in \eqref{QC} can be obtained from \eqref{abstract-QC} by specifying it to the adjoint representation. Nevertheless, these sets of constraints are only equivalent if such adjoint representation is faithful, otherwise one has to take into account that the linear dependence relations between the $24$ generators have to be supplemented with the vanishing conditions for some of them.}
\beq
\label{QC}
f_{\alpha R[MN} f_{\beta PQ]}{}^R = 0  \hspace{10mm},\hspace{10mm} \epsilon^{\alpha \beta} \, f_{\alpha MNR} \,f_{\beta PQ}{}^R = 0 \ . 
\eeq

The combination of supersymmetry and gaugings then induces the following scalar potential\footnote{We have set the gauge coupling constant to $\,g=\frac{1}{2}\,$ with respect to the conventions in ref.~\cite{Schon:2006kz}.}
\begin{align}
  V& = \dfrac{1}{64} \, f_{\alpha MNP} \, f_{\beta QRS} M^{\alpha
  \beta} \left[ \dfrac{1}{3} \, M^{MQ} \, M^{NR} \, M^{PS} + \left(\dfrac{2}{3} \, \eta^{MQ} - M^{MQ} \right) \eta^{NR}
  \eta^{PS} \right]  \notag \\
  & - \dfrac{1}{144} \, f_{\alpha MNP} \, f_{\beta QRS} \, \epsilon^{\alpha
  \beta} \, M^{MNPQRS}\, , \label{V}
 \end{align}
where
\beq 
\label{MV6}
M_{MNPQRS} \equiv \epsilon_{\underline{mnpqrs}}\mathcal{V}_{M}^{\phantom{M}\underline{m}}\mathcal{V}_{N}^{\phantom{M}\underline{n}}\mathcal{V}_{P}^{\phantom{M}\underline{p}}\mathcal{V}_{Q}^{\phantom{M}\underline{q}}\mathcal{V}_{R}^{\phantom{M}\underline{r}}\mathcal{V}_{S}^{\phantom{M}\underline{s}} \ .
\eeq
The underlined indices here are time-like rather than light-like, and related by the change of basis
\beq 
\label{lc_to_cart} 
R \equiv \frac{1}{\sqrt{2}} \left(
\begin{array}{cc}
 -\mathds{I}_{6} & \mathds{I}_{6} \\
  \phantom{-}\mathds{I}_{6} & \mathds{I}_{6}
\end{array}
\right) \ . 
\eeq
Because of this distinction between time- and space-like indices of SO$(6,6)$, this completely antisymmetric tensor is
invariant under local $\textrm{SO}(6) \times \textrm{SO}(6)$ transformations. Despite this, though, one would need to compute $\,\mathcal{V}\,$ associated with $\,M_{MN}\,$ explicitly in order to obtain the full form of the scalar potential.

\subsection{The SO($3$) truncation}

Let us consider the SO($3$) truncation of the full theory enjoying
an SL($2$) $\times$ SO($6,6$) global symmetry\footnote{This is the natural generalisation of the $\textrm{SL}(3) \times \textrm{SL}(3)$ truncation considered in ref.~\cite{Roest:2009dq}, and indeed will lead to a much richer landscape of vacua.}. In the following
sections of this work we will be dealing with (non-)geometric flux
compactifications of type II string theory having such a low-energy
effective description. This truncation is performed by considering
an SO($3$) subset in SO($6,6$) and keeping in the theory only the
singlets with respect to this subgroup both in the scalar sector and
in the embedding tensor part. Such a group theoretical
truncation is always guaranteed to be consistent in the sense that
all of the non-singlet scalars can be consistently set to zero in
that their field equations can never be sourced by SO($3$)
singlets. However, it by no means guarantees the stability of the non-singlets, and hence one must always explicitly check the mass spectrum of these fields as well.

\subsubsection*{The scalar sector of the theory}

The decomposition of the adjoint representation of $\textrm{SO}(6,6)$ contains six scalars
\beq
{\bf 66} \rightarrow 6 \cdot ({\bf 1},{\bf 1}) \oplus {\text{non-singlet representations}} \ ,
\eeq
amongst which two of them correspond to the product
$\textrm{SO}(6)\times \textrm{SO}(6)$ and therefore they are pure
gauge. This implies that the scalar coset associated with the matter
multiplets is parameterised in terms of only four physical scalars:
two dilatons ($\varphi_1$, $\varphi_2$) and two axions $(\chi_1,
\chi_2)$. The scalar coset in this sector reduces in the following
way under the SO(3) truncation
\beq
\frac{\textrm{SO}(2,2)}{\textrm{SO}(2)\times \textrm{SO}(2)} \ .
\eeq
The explicit parameterisation of $\,M_{M N}\,$ is defined in terms of a symmetric $\,G\,$ and an antisymmetric $\,B\,$ matrices as
\beq
\label{M_MN}
M_{M N} \equiv \left(
\begin{array}{cc}
G^{-1}  & - G^{-1} \, B \\
B \, G^{-1}    &  G - B\,G^{-1}\,B
\end{array}
\right) \ , 
\eeq
where $\,G\,$ and $\,B\,$ are given by
\beq 
\label{G&B} 
G= e^{\varphi_2-\varphi_1}\left(
\begin{array}{cc}
 \chi_2^2 + e^{-2\varphi_2} & -\chi_2\\
 -\chi_2 & 1
\end{array}
\right) \otimes \mathds{I}_{3}
\hspace{10mm} , \hspace{10mm}
B= \left(
\begin{array}{cc}
 0 & \chi_1 \\
 -\chi_1 & 0
\end{array}
\right) \otimes \mathds{I}_{3} \ .
\eeq
In consequence, we will choose the vielbein $\,\mathcal{V}\,$ in (\ref{M=VV}) to be
\beq 
\mathcal{V} \equiv \left(
\begin{array}{cc}
 \boldsymbol{e}^{T} & 0 \\
 B \, \boldsymbol{e}^{T} & \boldsymbol{e}^{-1}
\end{array}
\right) \otimes \mathds{I}_{3} 
\hspace{5mm} , \hspace{5mm} 
\boldsymbol{e} \equiv
e^{(\varphi_1+\varphi_2)/2} \left(
\begin{array}{cc}
 1 & \chi_2 \\
 0 & e^{-\varphi_2}
\end{array}
\right) \ ,
\eeq
with $\,\boldsymbol{e}^{T}\,\boldsymbol{e} = G^{-1}$. 

Using this parameterisation of the scalar sector in the truncated theory, the kinetic terms then reduce to
\beqa
\label{L_kin}
\mathcal{L}_{kin}&=&\dfrac{1}{8} \, (\partial M_{\alpha \beta})(\partial M^{\alpha \beta}) + \dfrac{1}{16} \, (\partial M_{MN})(\partial M^{MN}) \\[2mm]
&=& -\dfrac{1}{4}\left[(\partial\phi)^2+e^{2\phi}(\partial\chi)^2+3
(\partial\varphi_1)^2+3 \, e^{2\varphi_1}(\partial\chi_1)^2+3
(\partial\varphi_2)^2+3 \, e^{2\varphi_2}(\partial\chi_2)^2\right] \nonumber \ .
\eeqa

\subsubsection*{The quadratic constraints for the SO(3) truncation}

First of all, the number of allowed embedding tensor components
turns out to be 40, arranged into 20 SL(2) doublets, 20 being the
number of SO(3)-singlets contained in the decomposition of the ${\bf
220}$ of SO(6,6):
\begin{align}
  ({\bf 2, 220}) \rightarrow 20 \cdot ({\bf 2},{\bf 1}) \oplus {\text{non-singlet representations}}
  \, .
 \end{align}

\noindent A convenient way of describing these 20
$\textrm{SO}(3)$-invariant doublets is described in ref.~\cite{Derendinger:2004jn}, where the relevant components of the
embedding tensor are classified using the $\textrm{SO}(2,2)\times
\textrm{SO}(3)$ subgroup of $\textrm{SO}(6,6)$ with embedding ${\bf
12}=({\bf 4},{\bf 3})$. In this case, one can rewrite every
$\textrm{SO}(6,6)$ index $M$ as a pair $(A\,I)$, where $I=1,2,3$ is
a fundamental $\textrm{SO}(3)$ index, whereas $A=1,...,4$ is a
fundamental $\textrm{SO}(2,2)$ index. Due to this decomposition, the
structure constants of the gauge algebra can be factorised as
follows
\beq
\label{defL} 
f_{\alpha MNP} = f_{\alpha \, AI \, BJ \, CK}=\Lambda_{\alpha
ABC}\,\,\,\epsilon_{IJK},
\eeq
from which one can infer that the $\textrm{SO}(2,2)$-tensor
$\,\Lambda_{ABC}\,$ is \emph{completely symmetric}. This observation takes us
back to the number of 20 as expected from the group theoretical
decomposition. What one can now do, is to write down the quadratic
constraints \eqref{QC} in terms of the $\Lambda$ tensor. One obtains
\begin{align}
\epsilon^{\alpha\beta}\,\Lambda_{\alpha\,
AB}^{\phantom{ABC\,}C}\,\Lambda_{\beta DEC}=0\,,\qquad
\Lambda_{(\alpha\,
A[B}^{\phantom{ABC\,\,\,\,\,\,}C}\,\Lambda_{\beta)\, D]EC}=0\,,
\label{QCL}
\end{align}
where the extra indices $\alpha,\beta=(+,-)$ still represent the
SL$(2)$ phase.

The first set of constraints in \eqref{QCL} takes values in the
following representation of SL(2)$\times$ SO(2,2)
\Yvcentermath1\begin{align}\label{rep1} \left(\boldsymbol{1},
\tiny{\yng(2)\otimes_{AS}\yng(2)}\right)\,,
\end{align}
which has dimension 45, whereas the the second set of constraints in
\eqref{QCL} takes values in this other one
\Yvcentermath1\begin{align}\label{rep2}\left(\boldsymbol{3},
\tiny{\yng(2,2)}\,\right)\,,
\end{align}
which should not yet be thought of as only consisting of its
irreducible (traceless) part and therefore it has dimension 63. This
leads us to 108 as total amount of constraints, which can also be
obtained by means of a computer. It turns out, though, that the
number of independent constraints reduces to\footnote{This fact
should be understood in the following way: the trace part of
\eqref{rep2} is already implied by the remaining full set of
constraints coming from both \eqref{rep1} and \eqref{rep2}.} 105. We
will come back to this point in the next section when investigating the superpotential
formulation of our truncated theory.

\subsection{Relation to flux compactifications} 
\label{sec:N=1_form}

So far, we have introduced the main features of the SO($3$)
truncation of half-maximal supergravity in four dimensions. As we
have seen in the previous section, the scalar manifold in the
truncated theory reduces to 
\beq
\frac{\mathrm{SL}(2)}{\mathrm{SO}(2)}\,\times\,\frac{\mathrm{SO}(2,2)}{\mathrm{SO}(2)\,\times\,\mathrm{SO}(2)}\,\sim\,\left(\frac{\mathrm{SL}(2)}{\mathrm{SO}(2)}\right)^{3} \ , 
\eeq 
where each of the SL($2$) factors can be parameterised by a
complex scalar field. The resulting supergravity models are commonly
referred to in the literature as $STU$-models. They consist of
three complex fields which are related to those entering the
$\,M_{\alpha \beta}\,$ matrix in (\ref{chi,phi}) and the
$\,M_{MN}\,$ matrix in (\ref{M_MN}) -- through the metric $G$ and
the $B$-field in (\ref{G&B}) -- by
\beq
\label{complex_STU}
S \equiv \chi+i \, e^{-\phi}
\hspace{8mm},\hspace{8mm}
T \equiv \chi_1 + i \, e^{-\varphi_1}
\hspace{8mm} \textrm{and} \hspace{8mm}
U \equiv \chi_2 + i \, e^{-\varphi_2} \ .
\eeq
Furthermore, the splitting $\,\textbf{4}\rightarrow\,\textbf{1} \oplus
\textbf{3}\,$ of the fundamental representation of $\,\textrm{SU}(4)
\sim \textrm{SO}(6)\,$ R-symmetry under the action of
$\,\textrm{SO}(3)\,$ ensures an $\,\mathcal{N}=1\,$ structure of the
supergravity describing the truncated theory. This implies that it
has to be possible to formulate it in terms of a real K\"ahler
potential $\,K(\Phi,\bar{\Phi})\,$ and a holomorphic superpotential
$\,W(\Phi)\,$, where $\,\Phi=(S,T,U)\,$, by using the standard minimal supergravity formalism. According to it, the scalar potential can be worked out as
\beq
\label{scalar_potential}
V = e^K \left(  \sum_{\Phi} K^{\Phi\bar \Phi} |D_\Phi W|^2 - 3|W|^2 \right)  \ ,
\eeq
where $\,K^{\Phi \bar{\Phi}}\,$ denotes the inverse of the K\"ahler metric $\,K_{I\bar{J}} = \frac{\partial K}{\partial \Phi^{I} \partial \bar{\Phi}^{\bar{J}}}\,$, and $\,D_\Phi W = \frac{\partial W}{\partial \Phi} + \frac{\partial K}{\partial \Phi} W\,$ is the K\"ahler derivative.

\subsubsection*{The K\"ahler potential}

Let us start by noticing that the kinetic Lagrangian in \eqref{L_kin} can be rewritten in terms of the complex fields in (\ref{complex_STU}) as
\beq
\mathcal{L}_{kin}=K_{I\bar{J}}\,\partial\Phi^I\partial\bar{\Phi}^{\bar{J}}=\frac{\partial
S\partial \bar{S}}{\left(-i(S-\bar{S})\right)^2}+3 \, \frac{\partial
T\partial \bar{T}}{\left(-i(T-\bar{T})\right)^2}+3 \, \frac{\partial
U\partial \bar{U}}{\left(-i(U-\bar{U})\right)^2} \ ,
\eeq
with $\,K_{I\bar{J}}\,$ being again the K\"ahler metric. The above kinetic terms are then reproduced from the K\"ahler potential
\beq
\label{Kahler_pot}
K = - \,\log\left( -i\,(S-\bar{S})\right)  - 3 \,\log\left(-i\,(T-\bar{T})\right)  -3 \,\log\left( -i\,(U-\bar{U})\right) \ ,
\eeq
which matches the one obtained in string compactifications and being valid to first order in the string and the sigma model perturbative expansions.

\subsubsection*{The superpotential: flux backgrounds in terms of the embedding tensor}

Finding out the precise superpotential $\,W_{\textrm{SO}(3)}(\Phi)\,$ from which to reproduce the scalar potential in (\ref{V}) is certainly not an easy task. The reason why is that both scalar potentials, namely the one computed from the superpotential and that of (\ref{V}), do not have to perfectly match each other but they have to coincide up to the quadratic constraints in (\ref{QCL}).

\begin{figure}[h!]
\begin{center}
\scalebox{0.87}[0.87]{
\begin{tabular}{ccccc}
\includegraphics[scale=0.5,keepaspectratio=true]{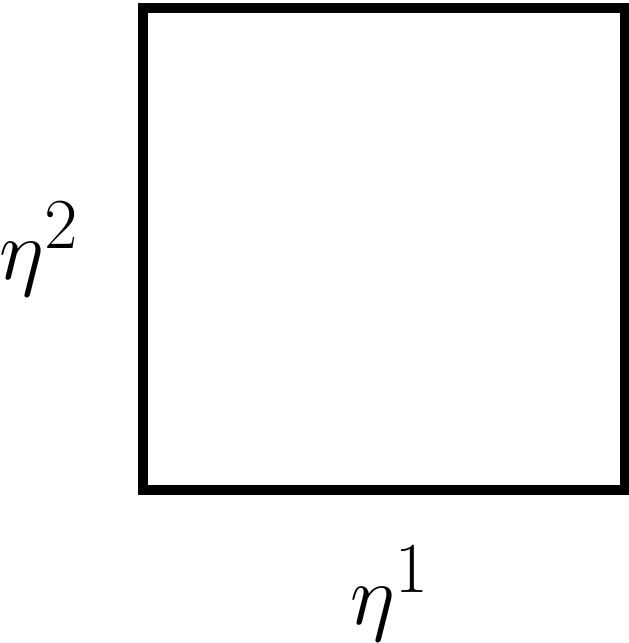} &    &  \includegraphics[scale=0.5,keepaspectratio=true]{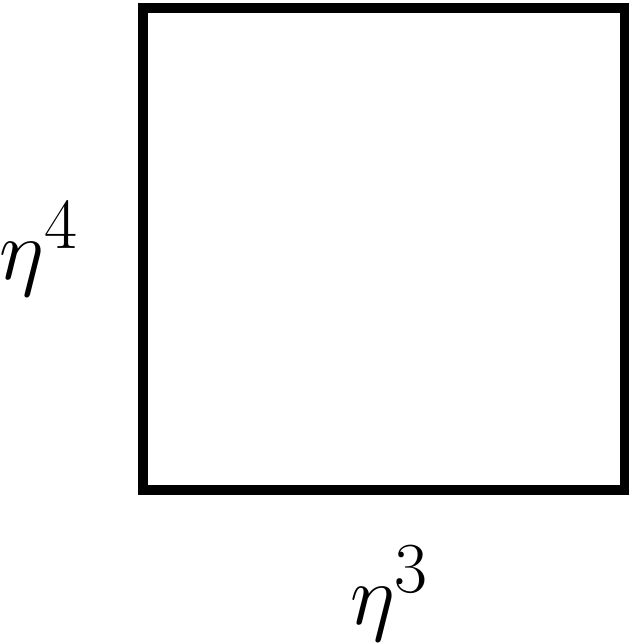}  &    & \includegraphics[scale=0.5,keepaspectratio=true]{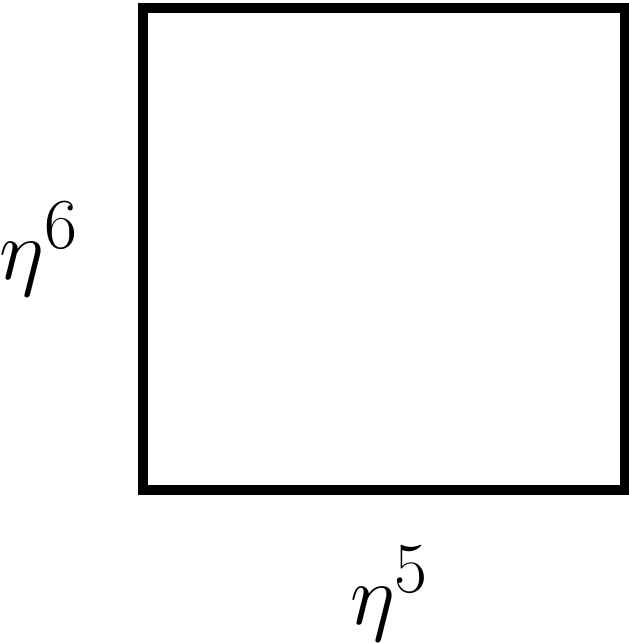} \\[-24mm]
     & \,\,\Large{$\times$} &        & \,\,\Large{$\times$} &
\end{tabular}
}
\end{center}
\vspace{9.5mm}
\caption{$\mathbb{T}^{6} =  \mathbb{T}^{2}_{1} \times \mathbb{T}_{2}^{2} \times \mathbb{T}_{3}^{2}$ torus factorisation and the coordinate basis.}
\label{fig:Torus_Factor}
\end{figure}

As for the above  K\"ahler potential, we want the superpotential $\,W_{\textrm{SO}(3)}(\Phi)\,$ also to stem from (orientifolds of) some string compactifications from ten to four dimensions. Their compatibility with producing an SO($3$) truncation of half-maximal supergravity in four dimensions allows for a simple interpretation of the internal space of the compactification. It can be taken to be the factorised six-torus of figure \ref{fig:Torus_Factor} whose coordinate basis is denoted $\eta^m$ with $m=1,\ldots,\,6\,$, supplemented with a set of flux objects fitting the embedding tensor components $\,f_{\pm MNP}\,$ surviving the truncation. 

\noindent
In the following we will use early Latin indices $a,b,c$ for horizontal $\,``-"$ $x$-like directions $(\eta^{1},\eta^{3},\eta^{5})$ and late Latin indices $i,j,k$ for vertical $\,``|"$ $y$-like directions $(\eta^{2},\eta^{4},\eta^{6})$ in the 2-tori $\,\mathbb{T}_{I}\,$ with $\,I=1,2,3$. This splitting of coordinates is in one-to-one correspondence with the SO($6,6$) index splitting of the embedding tensor components given in (\ref{defL}), where $A=(1,2,3,4) \equiv (a,i,\bar{a},\bar{i})\,$ refers to an SO($2,2$) fundamental index and $\epsilon_{IJK}$ denotes the usual totally antisymmetric tensor.

The identification between the embedding tensor components (gauging parameters) in the supergravity side and the flux objects in the string compactification side crucially depends on the string theory under investigation. As an example, when considering $\,\cN=1\,$ type IIA orientifold compactifications including O$6$-planes and D$6$-branes, only a few embedding tensor components in the supergravity side are known to correspond to flux components in the string theory side. In contrast, all of them correspond to (at least conjectured) fluxes in $\,\cN=1\,$ orientifold compactifications of type IIB string theory including O3/O7-planes and D3/D7-branes. In this type IIB scheme \cite{Aldazabal:2008zza,Dibitetto:2010rg}, the correspondence between embedding tensor components and fluxes entering the superpotential in (\ref{W_fluxes}) reads
\beq
\begin{array}{cccccccc}
f_{+mnp} = \tilde{F'}_{mnp} & \hspace{1mm},\hspace{1mm} & {f_{+\,mn}}^{p} = {{Q'}_{mn}}^{p} &  \hspace{1mm},\hspace{1mm}  &  f_{+\,\phantom{mn}p}^{\phantom{\a}mn\phantom{p}} = Q_{\,\phantom{mn}p}^{\,mn\phantom{p}} &  \hspace{1mm},\hspace{1mm}  &  f_{+}^{\phantom{+}mnp} = \tilde{F}^{mnp} & ,\\[2mm]
f_{-mnp} = \tilde{H'}_{mnp} & \hspace{1mm},\hspace{1mm} & {f_{-\,mn}}^{p} = {{P'}_{mn}}^{p} &  \hspace{1mm},\hspace{1mm}  &  f_{-\,\phantom{mn}p}^{\phantom{\a}mn\phantom{p}} = P_{\,\phantom{mn}p}^{\,mn\phantom{p}} &  \hspace{1mm},\hspace{1mm}  &  f_{-}^{\phantom{-}mnp} = \tilde{H}^{mnp} & ,
\end{array}
\eeq
where, for instance, $\tilde{F}^{mnp}\equiv \dfrac{1}{3!}\,\epsilon^{m \, n \, p \, m'n'p'} \, F_{m'n'p'}$. The correspondence between SO$(6,6)$ and SO$(2,2)$ embedding tensor components with known/conjectured flux objects in both type IIA and type IIB orientifold compactifications is presented in tables \ref{table:unprimed_fluxes} and \ref{table:primed_fluxes}.

\begin{table}[h!]
\renewcommand{\arraystretch}{1.25}
\begin{center}
\scalebox{0.87}[0.87]{
\begin{tabular}{ | c || c | c | c | c | c |}
\hline
couplings & SO($6,6$) & SO($2,2$) & Type IIB & Type IIA & fluxes\\
\hline
\hline
$1 $& $ -f_{+ \bar{a}\bar{b}\bar{c}} $  & $ - \Lambda_{+333} $& $ {F}_{ ijk} $& $F_{aibjck}$ & $  a_0 $\\
\hline
$U $& $f_{+ \bar{a}\bar{b}\bar{k}}$  &  $ \Lambda_{+334} $& ${F}_{ ij c} $& $F_{aibj}$ & $   a_1 $\\
\hline
$U^2 $& $ -f_{+ \bar{a}\bar{j}\bar{k}}$  & $ - \Lambda_{+344} $& ${F}_{i b c} $& $F_{ai}$ & $  a_2 $\\
\hline
$U^3 $& $f_{+ \bar{i}\bar{j}\bar{k}}$  & $ \Lambda_{+444} $& ${F}_{a b c} $& $F_{0}$ & $  a_3 $\\
\hline
\hline
$S $& $ -f_{- \bar{a}\bar{b}\bar{c}} $  &  $ - \Lambda_{-333} $& $ {H}_{ijk} $& $ {H}_{ijk} $  & $  b_0$\\
\hline
$S \, U $& $f_{- \bar{a}\bar{b}\bar{k}}$  & $ \Lambda_{-334}$ & $ {H}_{ij c} $& ${\omega}^{c}_{ij}$ & $  b_1 $\\
\hline
$S \, U^2 $& $ -f_{- \bar{a}\bar{j}\bar{k}}$   & $ - \Lambda_{-344} $& $ {H}_{ i b c}$ & $ {Q}_{ i }^{ b c}$  & $  b_2 $\\
\hline
$S \, U^3 $& $f_{- \bar{i}\bar{j}\bar{k}}$   &  $ \Lambda_{-444} $& $ {H}_{a b c} $& $ {R}^{a b c} $ & $  b_3 $\\
\hline
\hline
$T $& $f_{+ \bar{a}\bar{b}k}$   & $ \Lambda_{+233} $& $  Q^{a b}_k $&$ H_{a b k} $ & $  c_0 $\\
\hline
$T \, U $& $f_{+ \bar{a}\bar{j} k}=f_{+ \bar{i}\bar{b} k}\,\,\,,\,\,\,f_{+ a\bar{b}\bar{c}}$  &  $ \Lambda_{+234} \,\,\,,\,\,\, \Lambda_{+133} $& $ Q ^{a j}_k = Q^{i b}_k \,\,\,,\,\,\, Q^{b c}_a $& $ \omega^{j}_{k a} = \omega^{i}_{b k} \,\,\,,\,\,\, \omega_{b c}^a $  & $c_1 \,\,\,,\,\,\, \tilde {c}_1 $\\
\hline
$T \, U^2 $& $f_{+ \bar{i}\bar{b}c}=f_{+ \bar{a}\bar{j}c}\,\,\,,\,\,\,f_{+ \bar{i}\bar{j}k}$   &  $ \Lambda_{+134} \,\,\,,\,\,\, \Lambda_{+244} $& $ Q ^{ib}_c = Q^{a j}_c \,\,\,,\,\,\, Q^{ij}_k $& $ Q ^{ci}_b = Q^{j c}_a \,\,\,,\,\,\, Q^{ij}_k $ & $c_2 \,\,\,,\,\,\,\tilde{c}_2 $\\
\hline
$T \, U^3 $& $f_{+ \bar{i}\bar{j} c}$  &  $ \Lambda_{+144} $& $  Q^{ij}_{c} $& $  R^{ijc} $ & $c_3 $\\
\hline
\hline
$S \, T $& $f_{- \bar{a}\bar{b}k}$   &  $ \Lambda_{-233} $& $  P^{a b}_k $& & $  d_0 $\\
\hline
$S \, T \, U $& $f_{- \bar{a}\bar{j} k}=f_{- \bar{i}\bar{b} k}\,\,\,,\,\,\,f_{- a\bar{b}\bar{c}}$   &  $ \Lambda_{-234} \,\,\,,\,\,\, \Lambda_{-133} $& $ P ^{a j}_k = P^{i b}_k \,\,\,,\,\,\, P^{b c}_a $&  & $d_1 \,\,\,,\,\,\, \tilde {d}_1 $\\
\hline
$S \, T \, U^2 $& $f_{- \bar{i}\bar{b}c}=f_{- \bar{a}\bar{j}c}\,\,\,,\,\,\,f_{- \bar{i}\bar{j}k}$   &  $ \Lambda_{-134} \,\,\,,\,\,\, \Lambda_{-244} $& $ P ^{ib}_c = P^{a j}_c \,\,\,,\,\,\, P^{ij}_k $&  & $d_2 \,\,\,,\,\,\,\tilde{d}_2 $\\
\hline
$S \, T \, U^3 $& $f_{- \bar{i}\bar{j} c}$  &  $ \Lambda_{-144} $ & $  P^{ij}_{c} $&  & $d_3 $\\
\hline
\end{tabular}
}
\end{center}
\caption{Mapping between unprimed fluxes, embedding tensor components and couplings in the superpotential.}
\label{table:unprimed_fluxes}
\end{table}

\begin{table}[h!]
\renewcommand{\arraystretch}{1.25}
\begin{center}
\scalebox{0.87}[0.87]{
\begin{tabular}{ | c || c | c |c | c | c |}
\hline
couplings & SO($6,6$) & SO($2,2$) & Type IIB &  Type IIA & fluxes\\
\hline
\hline
$T^3 \, U^3 $& $ -f_{+ abc} $ & $ - \Lambda_{+111} $& $ {F'}^{ijk} $&  & $  a_0' $\\
\hline
$T^3 \, U^2 $&  $f_{+ abk}$ &  $ \Lambda_{+112} $& ${F'}^{ ij c} $& &$   a_1' $\\
\hline
$T^3 \, U $& $ -f_{+ ajk}$  & $ - \Lambda_{+122} $& ${F'}^{i b c} $& &$  a_2' $\\
\hline
$ T^3 $& $f_{+ ijk}$ & $ \Lambda_{+222} $& ${F'}^{a b c} $& &$  a_3' $\\
\hline
\hline
$S \, T^3 \, U^3 $& $ -f_{- abc} $  & $ - \Lambda_{-111} $& $ {H'}^{ ijk} $& &$  b_0'$\\
\hline
$S \, T^3 \, U^2 $& $f_{- abk}$  & $ \Lambda_{-112} $& $ {H'}^{i jc} $& &$  b_1' $\\
\hline
$S \, T^3 \, U $& $ -f_{- ajk}$  & $ - \Lambda_{-122} $& $ {H'}^{ i b c} $& & $  b_2' $\\
\hline
$S  \, T^3 $& $f_{- ijk}$   & $ \Lambda_{-222}$ & $ {H'}^{a b c} $& &$  b_3' $\\
\hline
\hline
$T^2 \, U^3 $& $f_{+ ab\bar{k}}$ & $ \Lambda_{+114} $& $  {Q'}_{a b}^k $& &$  c_0' $\\
\hline
$T^2 \, U^2 $& $f_{+ aj\bar{k}}=f_{+ ib\bar{k}}\,\,\,,\,\,\,f_{+ \bar{a}bc}$ & $  \Lambda_{+124} \,\,\,,\,\,\, \Lambda_{+113} $ & $ {Q'}_{a j}^k = {Q'}_{i b}^k \,\,\,,\,\,\, {Q'}_{b c}^a $& &$c_1' \,\,\,,\,\,\, \tilde{c}_1' $\\
\hline
$T^2 \, U $& $f_{+ ib\bar{c}}=f_{+ aj\bar{c}}\,\,\,,\,\,\,f_{+ ij\bar{k}}$  & $ \Lambda_{+123} \,\,\,,\,\,\, \Lambda_{+224} $& $ {Q'}_{ib}^c = {Q'}_{a j}^c \,\,\,,\,\,\, {Q'}_{ij}^k $& &$c_2' \,\,\,,\,\,\,\tilde{c}_2' $\\
\hline
$T^2 $& $f_{+ ij\bar{c}}$  & $ \Lambda_{+223} $& $  {Q'}_{ij}^{c} $& &$c_3' $\\
\hline
\hline
$S \, T^2 \, U^3$& $f_{- ab\bar{k}}$  & $ \Lambda_{-114} $& $  {P'}_{a b}^k $& &$  d_0' $\\
\hline
$S \, T^2 \, U^2 $& $f_{- aj\bar{k}}=f_{- ib\bar{k}}\,\,\,,\,\,\,f_{- \bar{a}bc}$ & $ \Lambda_{-124} \,\,\,,\,\,\, \Lambda_{-113} $& $ {P'}_{a j}^k = {P'}_{i b}^k \,\,\,,\,\,\, {P'}_{b c}^a $& &$d_1' \,\,\,,\,\,\, \tilde {d}_1' $\\
\hline
$S \, T^2 \, U $& $f_{- ib\bar{c}}=f_{- aj\bar{c}}\,\,\,,\,\,\,f_{- ij\bar{k}}$  &$ \Lambda_{-123} \,\,\,,\,\,\, \Lambda_{-224} $ & $ {P'}_{ib}^c = {P'}_{a j}^c \,\,\,,\,\,\, {P'}_{ij}^k $& &$d_2' \,\,\,,\,\,\,\tilde{d}_2' $\\
\hline
$S \, T^2  $& $f_{-ij\bar{c}}$ & $ \Lambda_{-223} $ & $  {P'}_{ij}^{c} $& &$d_3' $\\
\hline
\end{tabular}
}
\end{center}
\caption{Mapping between primed fluxes, embedding tensor components and couplings in the superpotential.}
\label{table:primed_fluxes}
\end{table}

Irrespective of the particular string theory realisation, we have explicitly checked that the scalar potential (\ref{V}) induced by the gaugings in the SO$(3)$ truncated theory is correctly reproduced, up to $\,\mathcal{N}=4\,$ quadratic constraints, from the following flux-induced superpotential 
\beq 
\label{W_fluxes} 
W_{\textrm{SO}(3)} = (P_{F} -
P_{H} \, S ) + 3 \, T \, (P_{Q} - P_{P} \, S ) + 3 \, T^2 \, (P_{Q'}
- P_{P'} \, S ) + T^3 \, (P_{F'} - P_{H'} \, S ) \ , 
\eeq
using the standard results in minimal supergravity. However, just by a simple inspection of tables \ref{table:unprimed_fluxes} and \ref{table:primed_fluxes}, it is clearly more convenient to adopt the terminology of the type IIB string theory when it comes to associate embedding tensor components to fluxes. In this picture, the superpotential in (\ref{W_fluxes}) contains flux-induced polynomials depending on both electric and magnetic pairs -- schematically $\,(e,m)\,$ -- of gauge $(F_{3},H_{3})$ fluxes and non-geometric $(Q,P)$ fluxes,
\beq
\begin{array}{lcll}
\label{Poly_unprim}
P_{F} = a_0 - 3 \, a_1 \, U + 3 \, a_2 \, U^2 - a_3 \, U^3 & \hspace{5mm},\hspace{5mm} & P_{H} = b_0 - 3 \, b_1 \, U + 3 \, b_2 \, U^2 - b_3 \, U^3 & ,  \\[2mm]
P_{Q} = c_0 + C_{1} \, U - C_{2} \, U^2 - c_3 \, U^3 & \hspace{5mm},\hspace{5mm} & P_{P} = d_0 + D_{1} \, U - D_{2} \, U^2 - d_3 \, U^3 & ,
\end{array}
\eeq
as well as those induced by their less known primed counterparts $\,(F'_{3},H'_{3})\,$ and $\,(Q',P')\,$ fluxes,
\beq
\begin{array}{lcll}
\label{Poly_prim}
P_{F'} = a_3' + 3 \, a_2' \, U + 3 \, a_1' \, U^2 + a_0' \, U^3 & \hspace{3mm},\hspace{3mm} &P_{H'} = b_3' + 3 \, b_2' \, U + 3 \, b_1' \, U^2 + b_0' \, U^3 & ,  \\[2mm]
P_{Q'} = -c_3' +  C'_{2} \, U + C'_{1} \, U^2 - c_0' \, U^3 & \hspace{3mm},\hspace{3mm} & P_{P'} = -d_3' + D'_{2} \, U + D'_{1} \, U^2 - d_0' \, U^3 & .
\end{array}
\eeq
For the sake of clarity, we have introduced the flux combinations $\,C_{i} \equiv 2 \, c_i - \tilde{c}_{i}\,$, $\,D_{i} \equiv 2 \, d_i - \tilde{d}_{i}\,$, $\,C'_{i} \equiv 2 \, c'_i - \tilde{c}'_{i}\,$ and $\,D'_{i} \equiv 2 \, d'_i - \tilde{d}'_{i}\,$ entering the superpotential, and hence the scalar potential and any other physical quantity.

These so-called primed fluxes have been conjectured in ref.~\cite{Aldazabal:2006up} to be needed in order to have a fully U-duality invariant flux background, but there is no further understanding of their physical role and of the types of sources coupling to them at the present stage. Still, those give a hint to understand the relation between doubled geometry and non-geometry as anticipated in the introduction. In the heterotic duality frame those two exactly coincide, in the sense that all the fluxes introduced by using doubled geometry happen to be interpretable as non-geometric fluxes. However, in such a duality frame it is impossible to introduce their magnetic dual counterparts. After performing an S-duality to go to type I (equivalent to type IIB with O9-planes) and subsequently a 6-tuple T-duality, we are in IIB with O3-planes. In such a duality frame, non-geometry and doubled geometry happen to give rise to two complementary generalised sets of fluxes, the second one consisting of these primed fluxes. Moreover, this particular frame is S-duality invariant and therefore such a flux background can be completed to a fully S-duality invariant one. This construction in the isotropic case allows us to at least formally\footnote{Primed fluxes do not have any well-defined string theory description, not even a local one, since they stem from some strongly coupled limit of the IIB theory.} describe all the embedding tensor components included in the SO($3$) truncation.

The superpotential in (\ref{W_fluxes}) was originally derived from a type II string theory approach in ref.~\cite{Aldazabal:2006up} by using duality arguments. Concretely, they worked out the $\,\cN=1\,$ duality invariant effective supergravity arising as the low energy limit of type II orientifold compactifications on the
$\,\mathbb{T}^{6}/(\mathbb{Z}_{2} \times \mathbb{Z}_{2})\,$ toroidal orbifold. More recently, this has been put in the context of type IIB (with O3/O7-planes)/F-theory compactifications in ref.~\cite{Aldazabal:2008zza} and connected to generalised geometry in ref.~\cite{Aldazabal:2010ef}. Finally, some aspects of the vacua
structure of this supergravity have been explored in refs~\cite{Font:2008vd,Guarino:2008ik,deCarlos:2009qm} where only the unprimed fluxes inducing the polynomials in (\ref{Poly_unprim}) were considered.

A worthwhile final remark about the SO$(3)$ truncation of half-maximal supergravity in four dimensions is that the resulting scalar potential $\,V\,$ is left invariant by the action of a discrete $\,\mathbb{Z}_{2}=\left\lbrace 1 \,,\,\alpha_{1} \right\rbrace\,$ symmetry. This parity symmetry transforms simultaneously the moduli fields $\,\Phi=(S,T,U)\,$ and the different fluxes $\,f_{i}\,$ as
\beq
\label{alpha1_sym}
\begin{array}{clccll}
 \alpha_{1} & : & \Phi   & \longrightarrow &  -\bar{\Phi} & , \\[2mm]
            &   &  f_{i} & \longrightarrow &  (-1)^{n_{1}+n_{2}+n_{3}} \, f_{i} & ,
\end{array}
\eeq
where $\,f_{i}\,\, S^{n_{1}} T^{n_{2}} U^{n_{3}}\,$ denotes a generic term in the superpotential (\ref{W_fluxes}). This transformation can be equivalently viewed as taking the superpotential from holomorphic to anti-holomorphic, i.e., $\,W(\Phi) \rightarrow$ $\,W(\bar{\Phi})$, without modifying the K\"ahler potential. This additional generator extends the SO$(2,2)$ part of the duality group to O$(2,2)$, while also acting with an element of determinant minus one on the SL$(2)$ indices.

\subsubsection*{Understanding the matching: are there unnecessary quadratic constraints?}

Let us go deeper into the matching between the $\,\mathcal{N}=1\,$ and $\cN = 4$ supergravity formulations of the theory. This
equivalence happens to hold only after the $\,\mathcal{N}=4\,$
quadratic constraints in (\ref{QCL}) are imposed on the
$\,\mathcal{N}=1\,$ side as well. Some of those constraints happen
to kill some moduli dependences which are not allowed by
$\,\mathcal{N}=4$, since they cannot be expressed in an
$\,\textrm{SL}(2) \times \textrm{SO}(6,6)\,$ covariant way, whereas
some others are only needed in order to recover the same
coefficients in front of terms which are present in both of the
theories. A further subtlety is that, in total, one only needs to
impose $96$ out of the $105$ independent quadratic constraints. This
means that there are 9 quadratic constraints which do not seem to be
needed in order for the matching to work. Going back to the
representation theory analysis we started in \eqref{rep1} and
\eqref{rep2}, one realises that \eqref{rep1} splits in the following
irreducible representations of SO($2,2$) in the case of the SO($3$)
truncated theory
\Yvcentermath1 \beq \label{rep3}
{\yng(2)\,\otimes_{\small{AS}}\,\yng(2)}={\yng(1,1)\,\oplus\,\yng(2)\,\oplus\,\yng(3,1)\,,}
\eeq
that is to say, a splitting of the $\boldsymbol{45}$ into
$\boldsymbol{6}\,\,\oplus\,\,\boldsymbol{9}\,\,\oplus\,\,\boldsymbol{30}$.
It turns out that all of the unneeded constraints combine together
to give the $\boldsymbol{9}$ irreducible component in the right-hand
side of \eqref{rep3}. The reason why these constraints are not
needed still remains unclear but it is a peculiar feature of the
SO($3$) truncation. This can be understood by going back to the full
theory, where those constraints combine together with other ones
into a bigger irreducible representation of $\,\textrm{SL}(2) \times
\textrm{SO}(6,6)\,$ and hence they have to be necessary as well as
the other constraints in order to have a complete matching between
the $\,\mathcal{N}=4\,$ and $\,\mathcal{N}=1\,$ scalar potentials.

Up to our knowledge, these results represent the first general demonstration\footnote{This point was also discussed in ref.~\cite{Aldazabal:2008zza} and we thank the authors for correspondence on their results.} of the explicit relation between the embedding tensor formulation of $\cN = 4$ supergravity and the superpotential formulation of $\cN =1$ supergravity in this particular truncation.

\section{Analysis of critical points}
\label{sec:crit}

In this section we present the strategy followed to find the
complete set of extrema of the scalar potential induced by the
gaugings and tools for analysing the mass spectrum and supersymmetry
breaking.
\subsection{Combining dualities and algebraic geometry techniques}
\label{sec:alg_geom}
The investigation of the full vacua structure of a
particular truncation is carried out by making use of the following
two ingredients: $i)$ part of the \,SL(2) $\times$ SO($2,2$)\, duality
group in order to reduce the extrema scanning to the origin of the
moduli space without loss of generality\footnote{This approach
differs from that followed in ref.~\cite{Font:2008vd} where the
invariance under the action of the duality group was used to remove
redundant flux configurations producing physically equivalent
solutions.}. $ii)$ specific algebraic geometry techniques which
permit an exhaustive identification of the flux backgrounds
producing such moduli solutions.
\\[1mm]
\noindent
Provided a set of vacuum expectation values (VEVs) for the moduli fields \mbox{$\Phi_{0}\equiv\left( S_{0},T_{0},U_{0}\right)$} that satisfies the extremisation conditions of the scalar potential, $\,\left.\partial_{\Phi}V \right|_{\Phi_{0}}=0\,$,\, it can always be brought to the origin of the moduli space, i.e.,
\beq
\label{moduli_origin}
S_{0}=T_{0}=U_{0}=i \ ,
\eeq
by subsequently applying a real shift together with rescaling upon each of the complex moduli fields. These transformations span the non-compact part, 
\beq
\label{Gnc}
G_{n.c} = \frac{\mathrm{SL}(2) \times \mathrm{SO}(2,2)}{\mathrm{SO}(2)^{3}} \ ,
\eeq
of the duality group. In the case of the modulus $\,S$, they belong to the electric-magnetic SL($2$) factor, while transformations on the moduli $T$ and $U$ belong to SO$(2,2)$. In consequence, the fluxes will also transform in such a way that they compensate the transformation of the moduli fields and leave the scalar potential invariant.

Because of the aforementioned, \textit{restricting the search of
extrema to the origin of the moduli space does not imply a lack of
generality as long as the considered set of flux components is
invariant under the action of the non-compact part of the duality group.}

This statement automatically leaves us with two complementary descriptions of the same problem: the field and the flux pictures. In the former, a consistent flux background is fixed and the problem reduces to the search of extrema of the scalar potential in the field space. In the latter, the point in field space is fixed (the origin) and the problem reduces to find the set of consistent flux backgrounds compatible with the origin being an extremum of the scalar potential.
\begin{figure}[h!]
\begin{center}
\scalebox{0.9}[0.9]{
\begin{tabular}{ccc}
\includegraphics[keepaspectratio=true]{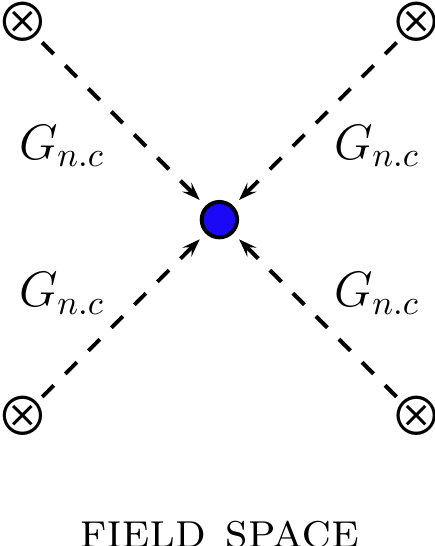} &  &  
 \hspace{15mm}\includegraphics[keepaspectratio=true]{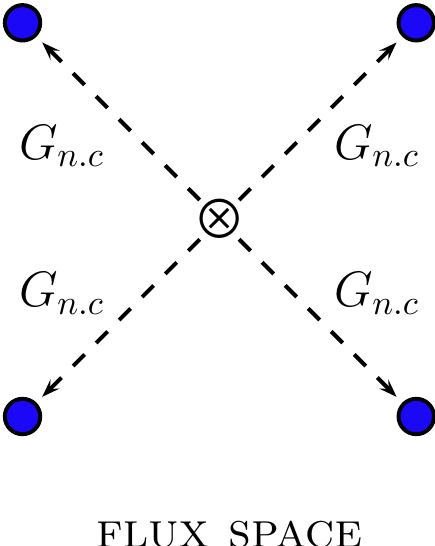} \\[-4cm]
&\hspace{15mm} \Large{$\longleftrightarrow$} &
\end{tabular}
}
\end{center}
\vspace{3cm}
\caption{Sketch of the correspondence between the field picture (crossed dots) and the flux picture (filled dots). The left diagram represents moduli space, whereas the right diagram illustrates the space of fluxes.}
\label{fig:field/flux_pic}
\end{figure}
The two descriptions are equivalent since dragging different moduli solutions down to the origin in the field space maps to a splitting of the corresponding flux background into various ones related by elements of $\,G_{n.c}\,$ in the flux space. This correspondence is depicted in figure~\ref{fig:field/flux_pic}.

Using the flux picture turns out to be quite useful because, schematically, the
scalar potential induced by the gaugings takes the form of
\beq V = \sum_{\textrm{terms}} (\textrm{\small{fluxes}})^{2}
\,\cdotp\, (\textrm{\small{fields}})^{\textrm{high degree}} \ , \eeq
hence being a sum of terms which are quadratic in the fluxes and
contain high degree couplings between the moduli fields. After
deriving the scalar potential with respect to the fields and going
to the origin of the moduli space, the extremum conditions reduce to
a set of quadratic constraints on the fluxes. Putting these
conditions together with the quadratic constraints in (\ref{QCL})
coming from the consistency of the gauging, we end up with a set of
homogeneous polynomial equations, namely an ideal $\,I\,$ in the
ring $\,\mathbb{C}\left[a_{0},\dots,d'_{3} \right]\,$, involving the
different flux components as variables,
\beq
\label{ideal_I}
I = \langle \,\left.\partial_{\Phi}V \right|_{\Phi_{0}} \,\,\, , \,\,\, \epsilon^{\alpha\beta}\,\Lambda_{\alpha\,
AB}^{\phantom{ABC\,}C}\,\Lambda_{\beta DEC} \,\,\, , \,\,\, \Lambda_{(\alpha\,
A[B}^{\phantom{ABC\,\,\,\,\,\,}C}\,\Lambda_{\beta)\, D]EC} \,  \rangle \ .
\eeq
Nonetheless, only those solutions for which all the flux components turn out to be real are physically acceptable.

The study of non-trivial multivariate polynomial systems and their link to geometry is the subject of algebraic geometry \cite{CLO_book}. A powerful computer algebra system for polynomial computations is provided by the \textsc{\,Singular\,} project \cite{DGPS}. Moreover, a comprehensive introduction to the specifics of this software as well as to the algebraic geometry techniques implemented on it can be found in ref.~\cite{GP_book}. These techniques have been shown to be a successful approach to investigate the vacua structure of the effective supergravity theories coming from flux compactifications of string theory \cite{Gray:2006gn,Guarino:2008ik} and some extensions including both fluxes and non-perturbative effects\footnote{For a computational implementation of these algebraic geometry tools into a \textit{Mathematica} package exploring vacuum configurations, see ref.~\cite{Gray:2008zs}.} \cite{Gray:2007yq}.

Among the set of algebraic geometry tools implemented within \textsc{\,Singular}, in this work we will make extensive use of the Gianni-Trager-Zacharias (GTZ) algorithm \cite{GTZ} for primary decomposition into prime ideals (for more details on primary decomposition algorithms, see the appendix B of ref.~\cite{Gray:2006gn} and references therein). Specifically, we will apply this method to decompose the ideal $I$ of (\ref{ideal_I}) into a set of $\,n\,$ simpler prime ideals $\,J_{n}\,$,
\beq
I= J_{1} \cap J_{2} \cap \ldots \cap J_{n} \ ,
\eeq
which can be solved analytically. These prime ideals will only intersect in a finite number of disjoint points and, in general, they may have different dimension.

For the sake of simplicity, we are not running this decomposition in
the most general case in which all the forty embedding tensor
components (fluxes) allowed in the SO($3$) truncation are kept.
Instead, we are considering two examples of gauged supergravities
which have a well understood interpretation as type II string
compactifications in the presence of flux backgrounds: type IIA
compactifications with gauge and metric fluxes
\cite{Derendinger:2004jn,Derendinger:2005ph,Aldazabal:2007sn,Dall'Agata:2009gv}
and type IIB compactifications with gauge fluxes
\cite{Giddings:2001yu,Kachru:2002he,DeWolfe:2004ns}.

Even though not all the fluxes are kept in these examples, the
previous argument for going to the origin of the moduli space
without loss of generality still holds since the transformation
needed to bring any moduli solution from its original location to
the origin (i.e. an element of $G_{n.c}$) does not turn on new flux components out of the initial
setup. We postpone a detailed analysis of more general flux
backgrounds for which a realisation in string theory is not known,
namely those including non-geometric fluxes, to future work.

\subsection{Supersymmetry breaking and full mass spectrum}
\label{sec:massBR}

Two further important steps in the analysis of critical points are
those of computing the amount of supersymmetry preserved at the
extrema of the $\,\mathcal{N}=4\,$ theory and the mass spectrum of
the scalar sector. As already pointed out in the introduction,
carrying out such a computation for a whole set of vacua can help us
shed further light on the relation between supersymmetry breaking
and instability, which has recently been a crucial point of
discussion in the context of extended supergravity. In order to do
this, we will compute the gravitini mass term included in the
fermionic mass terms Lagrangian \cite{Schon:2006kz}
\beq
\begin{array}{ccc}
e^{-1} \mathcal{L}_{\textrm{f.mass}} &\supset& \dfrac{1}{6} \,A_{1}^{ij} \, \bar{\psi}_{\mu i} \, \Gamma^{\mu \nu} \, \psi_{\nu
j} \ ,
\end{array}
\eeq
with $\,A_{1}^{ij}=A_{1}^{(ij)}$ and $\,i=1,...,4\,$ are SU($4$)
indices. This symmetric matrix is given in terms of the complexified
SL($2$) and SO($6,6$) vielbeins by
\beq
\label{A1_mass_matrix}
\begin{array}{ccc}
A_{1}^{ij} &=& \epsilon^{\alpha \beta} \, (\mathcal{V}_{\alpha})^{*}
\, {\mathcal{V}_{[kl]}}^{M} \, {\mathcal{V}_{N}}^{[ij]} \,
{\mathcal{V}_{P}}^{[jl]} \, {f_{\beta M}}^{NP} \ .
\end{array}
\eeq
The complexified SL($2$) vielbein $\,\mathcal{V}_{\alpha}\,$ is
written as
\beq \mathcal{V}_{\alpha} = e^{\phi/2} \, \left(
\bar{S} \,,\, 1 \right) \hspace{10mm} , \hspace{10mm} \textrm{where}
\hspace{5mm} S= \chi + i \, e^{-\phi} \ , \eeq
whereas the complexified (Lorentzian) SO($6,6$) vielbein
$\,{\mathcal{V}_{M}}^{[ij]}\,$ is built from the
$\,{\mathcal{V}_{M}}^{m}\,$ real vielbein by using the the mapping
\beq
\begin{array}{ccc}
\left( v^{12} \,,\, v^{13} \,,\, v^{14} \,,\, v^{34} \,,\, v^{42}
\,,\, v^{23} \right) &\equiv& \left( z_{1} \,,\, z_{2} \,,\, z_{3}
\,,\, z_{1}^{*} \,,\, z_{2}^{*} \,,\, z_{3}^{*} \right) \ ,
\end{array}
\eeq
where the complexification takes place as $\, z_{I} \equiv
\dfrac{1}{2} \left( v^{2I-1} + i \, v^{2I} \right) \,$ with
$\,\,I=1,2,3$. This is consistent with
\beq 
v_{ij} = (v^{ij})^{*} = \dfrac{1}{2} \, \epsilon_{ijkl} \, v^{kl} \ , 
\eeq
together with the normalisation
\beq 
- \, v^{m} \, \delta_{mn} \, v^{n} = - \, \dfrac{1}{2} \, \epsilon_{ijkl} \, v^{ij} \, v^{kl}  \ ,
\eeq
as was adopted in ref.~\cite{Schon:2006kz}. Using this matrix
$A_{1}^{ij}$, the Killing spinor equations determining the amount of
supersymmetry at any extremum is translated into the eigenvalues
equation
\beq 
\label{susy_cond}
A_{1}^{ij} \, q_{j} = \sqrt{-3 V_{0}} \, q^{i} \ ,
\eeq
where $\,q^{i}\,$ is an SU($4$) vector and $\,V_{0}\,$ is the
potential energy at either an AdS$_{4}$ or a Minkowski extremum.

Working in the SO($3$) truncation of the SO($6,6$) theory translates
into an $\,A_{1}^{ij}\,$ gravitini mass matrix of the general form
\beq 
A_{1} =  \textrm{diag}\left( \, \kappa_{1} \,\,\,,\,\,\, \kappa_{2} \,,\,\kappa_{2} \,,\,\kappa_{2} \, \right) 
\hspace{10mm} , \hspace{10mm} 
\textrm{with} \hspace{5mm} \kappa_{1},\kappa_{2} \in
\mathbb{C} \ , \label{kappa-eigenvalues}
\eeq
which reflects the splitting $\,\textbf{4}\rightarrow\,\textbf{1}
\oplus \textbf{3}\,$ of the fundamental of SU($4$) under the action
of SO($3$). Consequently one expects that the amount of
supersymmetry preserved would be
\begin{itemize}

\item[$i)$] $\mathcal{N}=4\,$ at those extrema where $\,|\kappa_{1}|=|\kappa_{2}|=\sqrt{-3 V_{0}}$.

\item[$ii)$] $\mathcal{N}=3\,$ at those extrema where $\,|\kappa_{1}| > |\kappa_{2}|\,$ with $\,|\kappa_{2}|=\sqrt{-3 V_{0}}$.

\item[$iii)$] $\mathcal{N}=1\,$ at those extrema where $\,|\kappa_{1}|< |\kappa_{2}|\,$ with $\,|\kappa_{1}|=\sqrt{-3 V_{0}}$.

\item[$iv)$] $\mathcal{N}=0\,$ at any other extremum.

\end{itemize}
The presented conditions
for preserving supersymmetry only constrain the modulus of the
eigenvalues of $A_1$ since the relation \eqref{susy_cond} exhibits a
U($1$) $\times$ U($1$) covariance. The action of these
transformations can be expressed in terms of the diagonal matrix $\text{diag}(\lambda\,, \,\mu\,,\,\mu\,,\,\mu)$, where $\lambda$, $\mu$ $\in$ U($1$).

Now it is worthwhile making a comment about the computation of the
full mass spectrum of the scalar sector for a vacuum of the
$\mathcal{N}=4\,$ theory. To this purpose we applied the mass
formula given in ref.~\cite{Borghese:2010ei}, where the scalar potential
of the full $\mathcal{N}=4\,$ theory has been expanded up to second
order around the origin in order to be able to read off the second
derivatives of the potential with respect to all of the 38 scalars
of the theory evaluated in the origin of moduli space. The Hessian
matrix evaluated in the origin is nevertheless not yet the physical
mass matrix from where one can draw conclusions about stability of a
solution. Suppose one has
\beq \label{Lag_can} e^{-1}
\mathcal{L}_{\textrm{canonic}}=\dfrac{1}{2} \,R -\dfrac{1}{2} \,
K_{ij} \, (\partial \phi^{i})(\partial \phi^{j}) - V \ ,\eeq
where $\,i=1,...,38\,$, then the covariant normalised mass$^2$ at an extremum
$\phi_0$ of the scalar potential $V$ is then given by
\beq 
\label{mass}
{(\textrm{mass}^{2})^{i}}_{j} =\left.  \frac{1}{|V|} \, K^{ik} \,\,
\dfrac{\partial^{2} V}{\partial \phi^{k} \partial \phi^{j}}
\right|_{\phi=\phi_{0}} \ , 
\eeq
where $\,K^{ij}\,$ denotes the inverse of the matrix $\,K_{ij}$
appearing in \eqref{Lag_can}. This (mass)$^{2}$ matrix is known as
the canonically normalised mass matrix, which is consistent with
taking the ``mostly plus'' signature for the space-time metric and
its eigenvalues are to be read as the values for the squared mass in
natural units\footnote{Every numerical value given in the following
sections for the energy is computed by setting the
reduced Planck mass $m_p$ to $1$, whereas one needs to reinsert the
value $\,m_{p}=\left( 8 \, \pi \, G \right)^{-1/2} = 2.43 \times 10^{18}
\,\, \textrm{GeV} \,$ when expressing quantities in energy units.}. According to this
definition of covariant mass, the Breitenlohner-Freedman (B.F.)
bound for the stability of an AdS$_4$ moduli solution is given by
\beq 
\label{BF_bound}
m^2\geq-\frac{3}{4}\ , 
\eeq 
where $m^2$ denotes the
lightest eigenvalue of the mass matrix \eqref{mass} at the AdS$_4$
extremum. The mass formulae for the masses of the SL($2$) scalars, those
ones of the SO($6,n$) sector and finally the mixing between them are given in ref.~\cite{
Borghese:2010ei}. In the next sections, when presenting results, we shall give
both a table with the values of the masses of the scalars in the
SO($3$) truncation and the full mass spectrum for comparison's sake.

\section{Geometric type IIA flux compactifications}
\label{sec:geom_IIA}

Let us commence this section by analysing the complete vacua structure of the SO$(3)$ truncation of $\,\cN=4$ supergravity which arises as the low energy limit of certain type IIA orientifold compactifications including background fluxes, D6-branes and O6-planes. More concretely, it is obtained from type IIA orientifold compactifications on a $\mathbb{T}^{6}/(\mathbb{Z}_{2} \times \mathbb{Z}_{2})$ isotropic orbifold in the presence of gauge Ramond-Ramond (R-R) ($F_{0}$, $F_{2}$, $F_{4}$, $F_{6}$) and Neveu-Schwarz-Neveu-Schwarz (NS-NS) $H_{3}$ fluxes, together with metric $\omega$ fluxes, D$6$-branes and O$6$-planes. In order to preserve half-maximal supersymmetry in four dimensions, the D$6$-branes have to be parallel to the O$6$-planes, i.e. they wrap the $3$-cycle in the internal manifold which is invariant under the action of the orientifold involution\footnote{Sources invariant under the combined action of the orientifold involution and the orbifold group break from half-maximal to minimal supersymmetry in four dimensions.}.

According to the mapping between fluxes and SO$(3)$-invariant embedding tensor components listed in table~\ref{table:unprimed_fluxes}, this type IIA flux compactification gives rise to an $\,\cN=4\,$ gauged supergravity for which the possible gaugings are determined in terms of the electric and magnetic flux parameters
\beq
\begin{array}{lclclclclc}
f_{+ \bar{a}\bar{b}\bar{c}} = -a_{0} & , & f_{+ \bar{a}\bar{b}\bar{k}}=a_{1}   &  ,  &  f_{+ \bar{a}\bar{j}\bar{k}}=-a_2  & , & f_{+ \bar{i}\bar{j}\bar{k}}=a_{3} & , \\[2mm]
f_{- \bar{a}\bar{b}\bar{c}}=-b_0 & , & f_{- \bar{a}\bar{b}\bar{k}}=b_{1} & , & f_{+ \bar{a}\bar{b}k} = c_{0} & , & f_{+ \bar{a}\bar{j} k}=f_{+ \bar{i}\bar{b} k}= c_{1} & , & f_{+ a\bar{b}\bar{c}}=\tilde{c}_{1} & .
\end{array}
\eeq
It is worth noticing here that in the type IIA scheme: $\,(a_{0}, \,a_{1}, \,a_{2}, \,a_{3})\,$ are R-R fluxes, $\,(b_{0}, \,c_{0})\,$ are NS-NS $\,H_{3}$-fluxes and $\,(b_{1}, \,c_{1}, \,\tilde{c}_{1})\,$ are metric $\,\omega$-fluxes. As we proved in the section~\ref{sec:N=1_form}, this effective supergravity admits an $\cN=1$ formulation in terms of the K\"ahler potential in (\ref{Kahler_pot}) and the superpotential
\beq
\label{W_IIA}
W_{\textrm{IIA}}=a_0 - 3 \, a_1 \, U + 3 \, a_2 \, U^2 - a_3 \, U^3 - b_0\,S + 3 \, b_1 \, S\,U + 3 \, c_0 \, T + (6 \, c_{1} - 3\, \tilde{c}_{1}) \, T\,U\ .
\eeq
Observe how acting upon this supergravity with the non-compact part of the duality group, i.e. rescalings and real shifts of the moduli fields, will not turn on new couplings in the superpotential (\ref{W_IIA}).

The quadratic constraints in (\ref{QC}) coming from the consistency of the $\,\cN=4\,$ gauging give rise to the three flux relations 
\beq
\label{QC_IIA}
c_{1}\,(c_{1}-\tilde{c}_{1})=0 \hspace{7mm} , \hspace{7mm} b_{1}\,(c_{1}-\tilde{c}_{1})=0 \hspace{7mm} , \hspace{7mm} - a_{3} \, c_{0} - a_{2} \, (2\, c_{1}-\tilde{c}_{1})=0 \ .
\eeq
The first and the second are respectively identified with the nilpotency ($d^{2}=\omega^{2}=0$) of the exterior derivative operator $d=\partial + \omega \, \wedge\,$ and the closure of the NS-NS flux background $\,dH_{3}=\omega \wedge H_{3}=0\,$. The third one is however related to the flux-induced tadpole
\beq
\label{Tad6}
\int_{10\textrm{d}} (\omega \wedge F_{2} + H_{3} \wedge F_{0}) \wedge C_{7}
\hspace{5mm} \Rightarrow \hspace{5mm}
N_{6}=\omega \wedge F_{2} + H_{3} \wedge F_{0} \ ,
\eeq
for the R-R gauge potential $\,C_{7}\,$ that couples to the D$6$-branes. In particular, it corresponds to the vanishing of the components along the internal directions orthogonal to the O$6$-planes,
\beq
\label{N6_orth}
N^{\bot}_{6}= - a_{3} \, c_{0} - a_{2} \, (2\, c_{1}-\tilde{c}_{1}) = 0 \ .
\eeq
In contrast, the component parallel to the O$6$-planes, denoted $\,N^{||}_{6}\,$, remains unrestricted since it can be canceled by adding sources still preserving half-maximal supersymmetry
\beq
N^{||}_{6}= 3 \, a_{2} \, b_{1} - a_{3} \, b_{0} \ .
\eeq
Nevertheless, whenever $\,N^{||}_{6}=0\,$ for a consistent flux background, then the resulting gauged supergravity admits an embedding into an $\,\cN=8\,$ theory. As a result, the flux background does not induce a tadpole for the $\,C_{7}$ gauge potential, i.e., $N^{\bot}_{6}=N^{||}_{6}=0$, and an enhanced four-elements discrete $\,\mathbb{Z}_{2} \times \mathbb{Z}_{2}=\left\lbrace 1 \,,\, \alpha_{1} \,,\, \alpha_{2} \,,\, \alpha_{1}\alpha_{2} \right\rbrace\,$ symmetry group shows up when it comes to relate non-equivalent vacuum configurations. 

This $\,\mathbb{Z}_{2} \times \mathbb{Z}_{2}\,$ discrete group is generated by the $\,\alpha_{1}$-transformation in (\ref{alpha1_sym}) and an extra parity transformation defined by
\beq
\label{alpha2_sym}
\begin{array}{clccll}
 \alpha_{2} & : & U   & \longrightarrow &  -\bar{U} & , \\[2mm]
            &   &  f_{i} & \longrightarrow &  (-1)^{n_{3}+1} \, f_{i} & ,
\end{array}
\eeq
where now $\,f_{i}\,\, S^{n_{1}} T^{n_{2}} U^{n_{3}}\,$ denotes a generic term in the superpotential of (\ref{W_IIA}). The action of the $\alpha_{2}$-transformation can equivalently be viewed as taking the original superpotential to a ``fake'' new one
\beq
\label{fake_W}
W_{\textrm{IIA}}(S,T,U) \rightarrow -W_{\textrm{IIA}}(S,T,\bar{U}) \ . 
\eeq
As a consequence, the scalar potential gets also modified as $\,V \rightarrow V + \delta V\,$ where $\,\delta V\,$ takes the form
\beq
\label{delta_V}
\delta V = \frac{1}{8\,(\textrm{Im}T)^{3}} \left[ \, 3 \, \left(\dfrac{\textrm{Im}T} {\textrm{Im}S}\right)\,N^{\bot}_{6} - N^{||}_{6} \, \right] \ .
\eeq
Therefore, having $N^{\bot}_{6}= N^{||}_{6} = 0$ (equivalently an $\,\cN=8\,$ flux background) ensures \mbox{$\delta V=0$} and hence a complete realisation of the $\,\mathbb{Z}_{2} \times \mathbb{Z}_{2}\,$ discrete group on the vacua distribution. The first $\,\mathbb{Z}_{2}\,$ factor relates a supersymmetric critical point to another supersymmetric one, while the second $\,\mathbb{Z}_{2}\,$ to a pair of fake supersymmetric critical points \cite{fake}.

The aim of this section is to completely map out the vacua structure of these $\,\cN=4\,$ type IIA compactifications. In particular, we are computing the complete set of extrema of the flux-induced scalar potential as well as the number of supersymmetries which they preserve and their mass spectrum. In the appendix~\ref{App:N1_vacua}, we have also studied the effect of introducing O$6$/D$6$ sources breaking from half-maximal to minimal supersymmetry, namely $\,N^{\bot}_{6}\neq 0\,$, and their consequences from the moduli stabilisation perspective.

\subsection{Full vacua analysis of the $\,\mathcal{N}=4\,$ theory}
\label{sec:extrema_N=4_IIA}

Here we will present the complete vacua data of the $\,\mathcal{N}=4\,$ supergravity theory introduced above. By this we mean to specify:
\begin{enumerate}
\item The complete set of vacua forming the landscape of the theory and the connections among themselves.

\item The associated data for each of these solutions: vacuum energy, supersymmetries preserved, mass spectrum and stability under fluctuations of all the scalar fields in the $\,\cN=4\,$ theory. 
\item The gauge group $\,G_{0}\,$ underlying the solutions.
\end{enumerate}

As it was explained in the previous section, algebraic geometry techniques are found to be powerful enough to find the entire set of extrema of the flux-induced scalar potential but, unfortunately, they will not give us any information about whether, and if so how, these extrema are linked to each other. To this respect, we will use the non-compact part $G_{n.c}$ of the duality group in (\ref{Gnc}) together with the discrete group generated by the transformations in (\ref{alpha1_sym}) and (\ref{alpha2_sym}) as an organising principle to connect different vacuum solutions. These connections will shed light upon the often confusing landscape of $\,\cN=4\,$ flux vacua. 

Our starting point is the ideal $\,I\,$ in (\ref{ideal_I}) consisting of the set of $\,\cN=4\,$ quadratic constraints in (\ref{QC_IIA}) together with the six extremisation conditions of the scalar potential with respect to the real and imaginary parts of the $S$, $T$ and $U$ fields evaluated at the origin of the moduli space. After decomposing it into prime factors, as explained in section~\ref{sec:crit}, we are left with a set of simpler pieces which can be solved analytically. The outcome of this process is a splitting of the landscape of vacua into sixteen pieces of dim$=1$ and an extra piece of dim$=2$. Let us go deeper into the features of these critical points.

\subsubsection*{The sixteen critical points of dim$=1$}

The sixteen critical points of $\,\dim=1\,$ in the $\,\cN=4\,$ theory are presented in table~\ref{table:N=4_vacua}. More concretely, we list the associated flux backgrounds after having brought these moduli solutions to the origin of the moduli space, as it was explained in detail in section~\ref{sec:alg_geom}. The vacuum energy at the solutions turns out to be
\beq
\label{V0_N=4}
V_{0}\left[ 1_{(s_1,s_2)} \right] = -\lambda^{2}
\hspace{3mm} , \hspace{3mm}
V_{0}\left[ 2_{(s_1,s_2)} \right] = V_{0}\left[ 4_{(s_1,s_2)} \right]= -\dfrac{32 \, \lambda^{2}}{27}
\hspace{3mm} , \hspace{3mm}
V_{0}\left[ 3_{(s_1,s_2)} \right] = -\dfrac{8 \, \lambda^{2}}{15}  \ .
\eeq

\begin{table}[t!]
\renewcommand{\arraystretch}{1.80}
\begin{center}
\scalebox{0.745}[0.745]{
\begin{tabular}{ | c || c | c |c | c | c | c |c | c |}
\hline
\textrm{\textsc{id}} & $a_{0}$ & $a_{1}$ & $a_{2}$ & $a_{3}$ & $b_{0}$ & $b_{1}$ & $c_{0}$ & $c_{1}=\tilde{c}_{1}$ \\[1mm]
\hline \hline
$1_{(s_1,s_2)}$ & $s_{2} \,  \dfrac{3 \,\sqrt{10}}{2}\, \lambda $ & $s_{1} \,\dfrac{\sqrt{6}}{2} \, \lambda$ & $ - s_{2} \,\dfrac{\sqrt{10}}{6} \, \lambda$ & $s_{1} \, \dfrac{5\,\sqrt{6}}{6} \, \lambda$ & $-s_{1} \,s_{2} \, \dfrac{\sqrt{6}}{3} \, \lambda$ & $\dfrac{\sqrt{10}}{3}\,\lambda$ & $s_{1} \,s_{2} \, \dfrac{\sqrt{6}}{3}\,\lambda$ & $\sqrt{10} \, \lambda$   \\[1mm]
\hline \hline
$2_{(s_1,s_2)}$ & $s_{2} \,\dfrac{16 \, \sqrt{10}}{9} \,\lambda$ & $0$ & $0$ & $s_{1} \, \dfrac{16 \, \sqrt{2}}{9} \, \lambda$ & $0$ & $\dfrac{16 \, \sqrt{10}}{45} \, \lambda$ & $0$ & $\dfrac{16 \, \sqrt{10}}{15} \, \lambda$   \\[1mm]
\hline
$3_{(s_1,s_2)}$ & $s_{2} \,\dfrac{4\,\sqrt{10}}{5}\,\lambda$ & $-s_{1} \, \dfrac{4\,\sqrt{30}}{15}\,\lambda$ & $s_{2} \, \dfrac{4\,\sqrt{10}}{15}\,\lambda$ & $s_{1} \,\dfrac{4\,\sqrt{30}}{15}\,\lambda$ & $s_{1} \, s_{2} \,\dfrac{4\,\sqrt{30}}{15}\,\lambda$ & $\dfrac{4\,\sqrt{10}}{15}\,\lambda$ & $-s_{1} \, s_{2} \,\dfrac{4\,\sqrt{30}}{15}\,\lambda$ & $\dfrac{4\,\sqrt{10}}{5}\,\lambda$   \\[1mm]
\hline
$4_{(s_1,s_2)}$ & $s_{2} \,\dfrac{16 \, \sqrt{10}}{9} \,\lambda$ & $0$ & $0$ & $s_{1} \,\dfrac{16 \, \sqrt{2}}{9} \,\lambda$ & $0$ & $\dfrac{16 \, \sqrt{2}}{9} \,\lambda$ & $0$ & $\dfrac{16 \, \sqrt{2}}{9} \,\lambda$  \\
\hline
\end{tabular}
}
\end{center}
\caption{The sixteen extrema of $\dim=1$ in the scalar potential of the $\mathcal{N}=4$ theory. They can be arranged into four groups of extrema each of which in turn consists of four solutions labelled by a choice of the pair of signs $(s_{1},s_{2})\equiv\left\lbrace (+,+) , (+,-) , (-,+) , (-,-) \right\rbrace$.} 
\label{table:N=4_vacua}
\end{table}

As we discussed in section~\ref{sec:massBR}, the number of supersymmetries preserved in these solutions can be computed from the gravitini mass matrix $\,A^{ij}_{1}\,$ in (\ref{kappa-eigenvalues}). After solving the eigenvalues equation of (\ref{susy_cond}), we find that all the solutions of the $\,\mathcal{N}=4\,$ theory are non-supersymmetric except those ones labelled by $\,1_{(+,+)}\,$ and $\,1_{(-,+)}\,$ which turn out to preserve $\mathcal{N}=1$ supersymmetry. Nevertheless, it is worth noticing here that they all actually enjoy an embedding in an $\,\mathcal{N}=8\,$ theory due to the lack of flux-induced tadpoles for the local sources\footnote{The condition $\,N_{6}^{||}=0\,$ is in fact implied by the $\mathcal{N}=4$ quadratic constraints and two of the three axionic field equations provided $c_0\,a_1\neq 0$. This is the case for the solutions $1_{(s_1,s_2)}$ and $3_{(s_1,s_2)}$ in table~\ref{table:N=4_vacua}, whereas for the flux background in the remaining cases it is straightforward.}, i.e.,  
\beq
N_{6}^{\bot} = N_{6}^{||} =  0 \ .
\eeq
This observation was previously made for the $\,\cN=1\,$ type IIA supersymmetric solution found in ref.~\cite{Dall'Agata:2009gv}. Now we are extending the statement about the existence of an $\,\cN=8\,$ lifting to the complete vacuum structure of the theory including both minimally supersymmetric and non-supersymmetric solutions. This fact has two immediate implications, the second actually being a direct consequence of the first:
\begin{itemize}
\item[$i)$] The discrete $\,\mathbb{Z}_{2}\,$ group generated by the $\,\alpha_{2}$-transformation in (\ref{alpha2_sym}) is ``accidentally'' realised as a symmetry of the flux-induced scalar potential $V(\Phi)$.  Then a complete discrete symmetry group $\,\mathbb{Z}_{2} \times \mathbb{Z}_{2}=\left\lbrace 1 \,,\, \alpha_{1} \,,\, \alpha_{2} \,,\, \alpha_{1}\alpha_{2} \right\rbrace\,$ appears in the landscape of the $\,\cN=4\,$ theory connecting solutions through the chain
\beq
\label{disc_chain}
N_{(+,+)} \,\,\overset{\alpha_{1}}{\longrightarrow}\,\, 
N_{(-,+)} \,\,\overset{\alpha_{2}}{\longrightarrow}\,\, 
N_{(-,-)} \,\,\overset{\alpha_{1}}{\longrightarrow}\,\, 
N_{(+,-)} \,\,\overset{\alpha_{2}}{\longrightarrow}\,\, 
N_{(+,+)} \ ,
\eeq
where $\,N=1,2,3,4\,$ stands for the four groups of solutions $\,N_{(s_{1},s_{2})}\,$ in table~\ref{table:N=4_vacua}. In fact, we have checked that combining these discrete transformations with the continuous non-compact part $\,G_{n.c}\,$ in (\ref{Gnc}) of the duality group, the vacua structure of the theory turns out to be a net of extrema connected by elements of the enhanced group
\beq
\label{Gvac}
G_{vac} = G_{n.c} \times \mathbb{Z}_{2} \times \mathbb{Z}_{2} \ .
\eeq
As it is shown in figure~\ref{fig:net}, all the sixteen critical points of $\,\dim=1\,$ in the $\,\cN=4\,$ theory are then connected to each other by an element of $G_{vac}$.

\begin{figure}[h!]
\vspace{5mm}
\begin{center}
\scalebox{0.7}[0.7]{
\includegraphics[keepaspectratio=true]{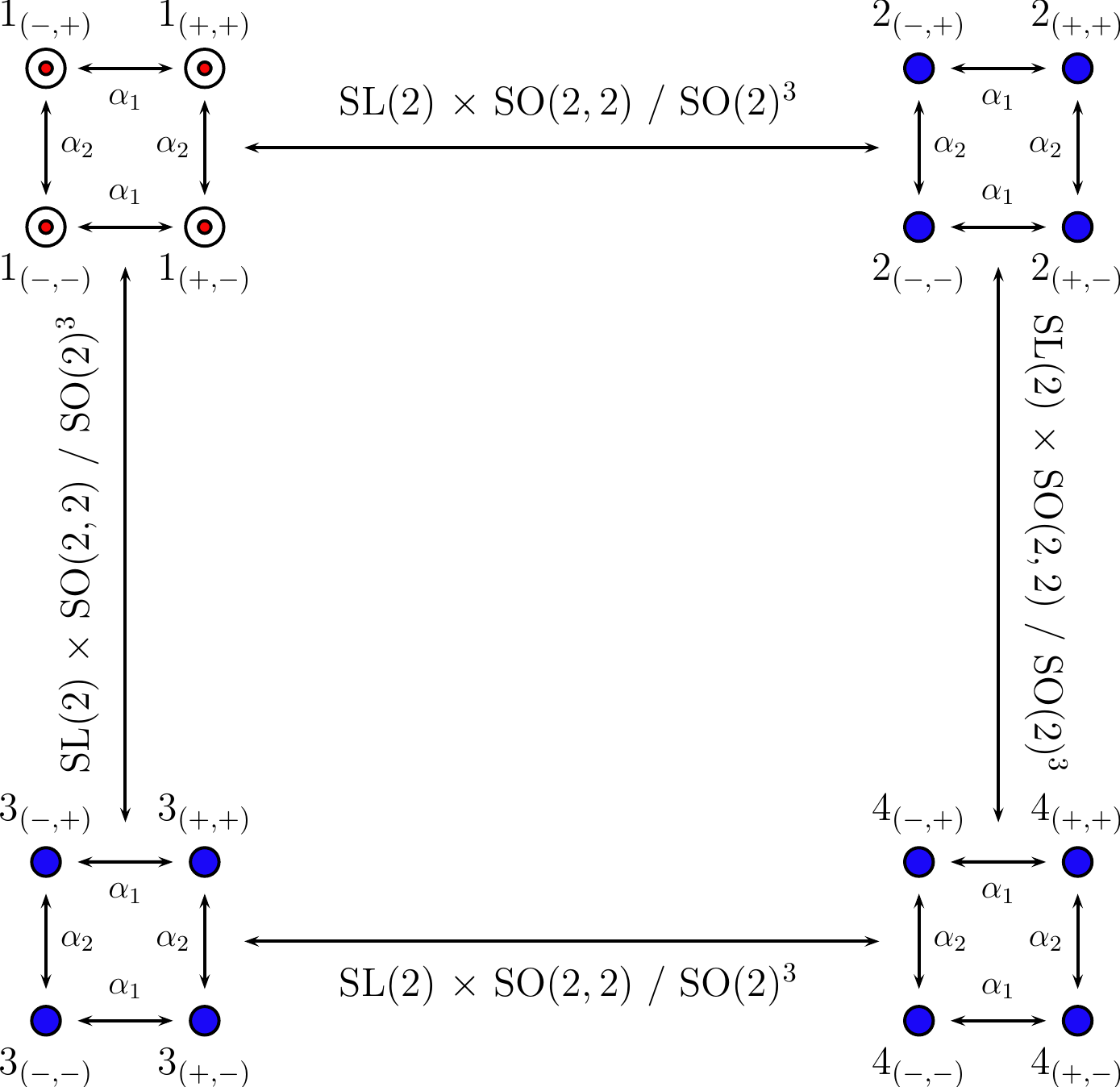} 
}
\end{center}
\caption{Net of connections between the $\,\dim=1\,$ sixteen critical points of the $\,\cN=4\,$ theory. The dotted points correspond to (fake-)supersymmetric solutions whereas the filled ones are non-supersymmetric.}
\label{fig:net}
\end{figure}

\item[$ii)$] Since the $\,\alpha_{2}$-transformation in (\ref{alpha2_sym}) is an accidental symmetry of the scalar potential but not of the superpotential, then the existence of non-supersymmetric and nevertheless stable solutions is guaranteed as long as there are supersymmetric ones. The reason is that these non-supersymmetric solutions would be ``fake'' supersymmetric in the sense that they do correspond to supersymmetric solutions of the ``fake'' superpotential in (\ref{fake_W}). Consequently, all the results concerning stability of supersymmetric solutions still apply to these non-supersymmetric ones since the scalar potential is left invariant. Supersymmetric and ``fake'' supersymmetric (non-supersymmetric) solutions of the theory are then connected by  
\beq
\label{susy/fake_chain}
\begin{array}{ccccccc}
\textrm{\footnotesize{SUSY}}&  & \textrm{\footnotesize{SUSY}}  &  & \textrm{\footnotesize{FAKE SUSY}}  &  &  \textrm{\footnotesize{FAKE SUSY}} \\
1_{(+,+)} &\overset{\alpha_{1}}{\longrightarrow}& 
1_{(-,+)} &\overset{\alpha_{2}}{\longrightarrow}& 
1_{(-,-)} &\overset{\alpha_{1}}{\longrightarrow}& 
1_{(+,-)}  
\end{array}
\,\,\,  \ . 
\nonumber
\eeq
We will see this explicitly by computing the full mass spectrum associated to these solutions and checking that they coincide.
\end{itemize}

The first step to check stability involves computing the masses only for the SO$(3)$-invariant fields, namely the $\,\textrm{SL}(2)/\textrm{SO}(2)\,$ axiodilaton $\,S\,$ and the two $\,\textrm{SO}(2,2)/\textrm{SO}(2)^{2}\,$ moduli fields $\,T\,$ and $\,U$. Nonetheless, stability of a solution under fluctuations of these $\,2+4=6\,$ real fields does not imply stability with respect to the rest of the $\,\mathcal{N}=4\,$ scalars which may render it unstable. The set of normalised masses of the SO$(3)$-invariant scalars at the sixteen $\,\dim=1\,$ extrema of the $\,\mathcal{N}=4\,$ theory are summarised in table~\ref{table:N=4_masses}. As we anticipated, they do not depend on the choice of a particular $\,(s_{1},s_{2})\,$ solution within a $\,N_{(s_{1},s_{2})\,}$ group. 

\begin{table}[h!]
\renewcommand{\arraystretch}{1.80}
\begin{center}
\scalebox{0.87}[0.87]{
\begin{tabular}{ | c || c | c | c | c | c | c | c |}
\hline
\textrm{\textsc{id}} & $m_{1}^{2}$ & $m_{2}^{2}$ & $m_{3}^{2}$ & $m_{4}^{2}$ & $m_{5}^{2}$ & $m_{6}^{2}$ & B.F. \\
\hline \hline
$1_{(s_1,s_2)}$ & $0$ & $-\dfrac{2}{3}$ & $\dfrac{4 + \sqrt{6}}{3}$ & $\dfrac{4 - \sqrt{6}}{3}$ & $\dfrac{47 + \sqrt{159}}{9}$ & $\dfrac{47 - \sqrt{159}}{9}$ & $m^{2}=-\dfrac{2}{3} \rightarrow \textrm{stable}$ \\[1mm]
\hline\hline
$2_{(s_1,s_2)}$ & $0$ & $-\dfrac{4}{5}$ & $-\dfrac{2}{5}$ & $2$ & $\dfrac{64}{15}$ & $\dfrac{20}{3}$ & $m^{2}=-\dfrac{4}{5} \rightarrow \textrm{unstable}$ \\[1mm]
\hline
$3_{(s_1,s_2)}$ & $0$ & $0$ & $2$ & $2$ & $\dfrac{20}{3}$ & $\dfrac{20}{3}$ & $\min$ \\[1mm]
\hline
$4_{(s_1,s_2)}$ & $0$ & $0$ & $\dfrac{4}{3}$ & $2$ & $6$ & $\dfrac{20}{3}$ & $\min$ \\[1mm]
\hline
\end{tabular}
}
\end{center}
\caption{Eigenvalues of the SO($3$)-truncated canonically normalised mass matrix at the AdS$_{4}$ extrema of the scalar potential in the $\mathcal{N}=4$ theory. For those being saddle points, the last column shows their stability according to the Breitenlohner-Freedman bound in (\ref{BF_bound}).} 
\label{table:N=4_masses}
\end{table}

Up to this point, the given information about the mass spectrum and stability of solutions is still incomplete. In order to determine whether these critical points are actually stable under fluctuations of all the scalar fields in the $\,\mathcal{N}=4\,$ theory, we have to compute the full mass spectrum. As already anticipated in section~\ref{sec:massBR}, we have made use of the mass formula provided in ref.~\cite{Borghese:2010ei} to address the issue of stability. The computation of the complete mass spectrum for the sixteen $\,\dim=1\,$ solutions of the $\,\mathcal{N}=4\,$ geometric type IIA compactifications gives the following results:

\begin{itemize}
\item The normalised scalar field masses and their multiplicities for the four solutions $\,1_{(s_{1},s_{2})}\,$ take the values of
\beq
\begin{array}{lcrr}
\dfrac{1}{9} \left(47 \pm \sqrt{159}\right)\,\,(\times 1)
\hspace{8mm} , \hspace{8mm} \dfrac{1}{3} \left(4 \pm
\sqrt{6}\right)\,\,(\times 1) & , &
\dfrac{29}{9}\,\,(\times 3) & ,\\[4mm]
\dfrac{1}{18} \left( \, 89 + 5 \, \sqrt{145} \pm \sqrt{606 + 30 \, \sqrt{145}} \, \right)\,\,(\times 5) 
& \hspace{5mm} , \hspace{5mm} &
0\,\,(\times 10) & , \\[4mm]
\dfrac{1}{18} \left(\, 89 - 5 \, \sqrt{145} \pm \sqrt{606 - 30 \, \sqrt{145}} \, \right)\,\,(\times 5)  
& \hspace{5mm} ,\hspace{5mm} &
-\dfrac{2}{3}\,\,(\times 1) & .
\end{array}
\nonumber 
\eeq
The unique tachyonic scalar then implies $\,m^2= - \tfrac23\,$ so these AdS$_4$ solutions satisfy the B.F. bound in (\ref{BF_bound}) hence being totally stable. Notice that the dangerous tachyonic mode has a special mass value, corresponding to a massless supermultiplet and being identical to that of a conformally coupled scalar field in AdS$_4$ \cite{Townsend}. In terms of group theory, it corresponds to the discrete unitary irreducible representation for AdS$_4$, while all other masses with $m^2 \geq - \tfrac34$ comprise a continuous family of such irreps. 

\item The normalised scalar field masses and their multiplicities for the four solutions $\,2_{(s_{1},s_{2})}\,$ take the values of
\beq
\begin{array}{c}
\dfrac{1}{15} \left(77 \pm 5 \, \sqrt{145}\right)\,\,(\times
5) \hspace{4mm} , \hspace{4mm} \dfrac{2}{15} \left(31 \pm
\sqrt{145}\right)\,\,(\times 5) \hspace{4mm} , \hspace{4mm}
\dfrac{64}{15}\,\,(\times 1) \hspace{4mm} , \hspace{4mm}
\dfrac{20}{3}\,\,(\times 1) \ , \\[4mm]
\dfrac{46}{15}\,\,(\times 3) \hspace{5mm} , \hspace{5mm}
2\,\,(\times 1) \hspace{5mm} , \hspace{5mm}
0\,\,(\times 10) \hspace{5mm} , \hspace{5mm}
-\dfrac{2}{5}\,\,(\times 1) \hspace{5mm} , \hspace{5mm}
-\dfrac{4}{5}\,\,(\times 1) \ .
\end{array}
\nonumber 
\eeq
In this case the most tachyonic mode gives rise to $\,m^2=-4/5\,$ that is below the B.F. bound in (\ref{BF_bound}), so these AdS$_4$ solutions become unstable under fluctuations of this mode.

\item The normalised scalar field masses and their multiplicities for the four solutions $\,3_{(s_{1},s_{2})}\,$ take the values of
\beq 
\frac{1}{3} \left(19\pm\sqrt{145}\right)\,\,(\times
10) \hspace{4mm} , \hspace{4mm} \frac{20}{3}\,\,(\times 2)
\hspace{4mm} , \hspace{4mm} \frac{14}{3}\,\,(\times 3)
\hspace{4mm} , \hspace{4mm} 2\,\,(\times 2)
\hspace{4mm} , \hspace{4mm} 0\,\,(\times 11) \ , \nonumber 
\eeq
whereas those corresponding to the four solutions $\,4_{(s_{1},s_{2})}\,$ are given by  
\beq
\dfrac{20}{3}\,\,(\times 1) \hspace{4mm} , \hspace{4mm}
6\,\,(\times 6) \hspace{4mm} , \hspace{4mm}
\dfrac{8}{3}\,\,(\times 5) \hspace{4mm} , \hspace{4mm}
2\,\,(\times 4) \hspace{4mm} , \hspace{4mm}
\dfrac{4}{3}\,\,(\times 6) \hspace{4mm} , \hspace{4mm}
0\,\,(\times 16) \,\ .
\nonumber 
\eeq
One observes that all the normalised masses are non-negative so these AdS$_{4}$ solutions do actually correspond to stable extrema of the scalar potential.
\end{itemize}
Therefore, this shows that most of the AdS$_{4}$ moduli solutions of the $\,\mathcal{N}=4\,$ theories coming from geometric type IIA flux compactifications are non-supersymmetric and nevertheless stable even when considering all the $\,2+36=38\,$ scalar fields\footnote{It would be interesting to understand the (dis-)similarities with the non-supersymmetric vacua in refs~\cite{Lust:2008zd,Koerber:2010rn}.}. 

A point to be highlighted is that, in this type IIA case, the SO($3$) truncation turns out to capture the interesting dynamics of the scalars, in the sense that the lightest mode is always kept by the truncation. This is by no means guaranteed by the consistency of the truncation. Indeed, as was discussed in the introduction, there are $\cN = 8$ examples of consistent truncations where the non-singlets lead to instabilities of critical points that are stable with respect to the singlet sector \cite{Warner:1006}. The situation for the critical points here differs from this in two respects. Firstly, the non-singlet masses always lie above the lightest mode in the singlet sector. Moreover, the non-singlet masses are in fact always non-negative.

Another remarkable feature is that the supersymmetric solutions $\,1_{(+,+)}\,$ and $\,1_{(-,+)}\,$ are not the (stable) ones with highest potential energy. Indeed, the solutions $\,3_{(s_{1},s_{2})}\,$ are non-supersymmetric and still stable with a higher vacuum energy, as can be read from (\ref{V0_N=4}).  This again differs from the situation in the prototypical $\cN = 8$ supergravity with SO$(8)$ gauging, where the vacuum that preserves all supersymmetry has the highest potential energy of all known critical points \cite{Fischbacher:2009cj}.

Finally we want to identify the gauge group(s) $\,G_{0}\,$ underlying these solutions. The antisymmetry of the brackets in (\ref{SW_algebra_ZX}), when restricted to the fluxes compatible with type IIA geometric backgrounds, allows to write the magnetic generators in terms of the electric ones
\beq
X_{-}^{a} = - \frac{(b_1\,c_0 + b_0\,c_1)}{c_1 \, \tilde{c}_{1}} \,\, Z_{+a} \,+\, \frac{b_{1}}{\tilde{c}_{1}} \,\, Z_{+i}
\hspace{5mm} , \hspace{5mm}
X_{-}^{i} = \frac{b_1}{c_1} \,\, Z_{+a}
\hspace{5mm} , \hspace{5mm}
Z_{-a}=Z_{-i} = 0 \ ,
\eeq
with pairs $\,(a,i)=\left\lbrace (1,2),(3,4),(5,6) \right\rbrace\,$. Notice that $\,c_1 \, \tilde{c}_{1} \neq 0\,$ for all the solutions listed in table~\ref{table:N=4_vacua}. In terms of electric generators, the algebra $\,\mathfrak{g_{0}}\,$ of $\,G_{0}\,$ is expressed as a twelve dimensional algebra which is now suitable to define a consistent gauging of the theory. The brackets involving isometry-isometry generators are given by
\beq
[Z_{+a},Z_{+b}]=[Z_{+a},Z_{+j}]=[Z_{+i},Z_{+j}]=0 \ ,
\eeq
and then span an abelian $\,\mathfrak{u}(1)^{6\,}$ subalgebra of $\,\mathfrak{g_{0}}$. Furthermore, the mixed non-vanishing isometry-gauge brackets read
\beq
[Z_{+a},X_{+}^{b}] = \tilde{c}_{1}\,Z_{+c}
\hspace{5mm} , \hspace{5mm}
[Z_{+i},X_{+}^{b}] = c_{0}\,Z_{+c} \,+\, c_{1} \, Z_{+k}
\hspace{5mm} , \hspace{5mm}
[Z_{+i},X_{+}^{j}] = c_{1} \, Z_{+c} \ ,
\eeq
so the isometry generators actually determine an abelian ideal within $\,\mathfrak{g_{0}}$. Accordingly to the Levi's decomposition theorem, the algebra $\,\mathfrak{g_{0}}\,$ can then be written as
\beq
\mathfrak{g_{0}} = \mathfrak{g}_{\textrm{gauge}} \oplus \mathfrak{u(1)^{6}} \ ,
\eeq
where $\,\mathfrak{g}_{\textrm{gauge}}\,$ has to be read off from the gauge-gauge brackets after quotienting $\,\mathfrak{g_{0}}\,$ by the abelian ideal. They take the form of
\beq
[X_{+}^{a},X_{+}^{b}] = \tilde{c}_{1} \, X_{+}^{c} + c_{0}\,X_{+}^{k}
\hspace{5mm} , \hspace{5mm}
[X_{+}^{a},X_{+}^{j}] = c_{1}\,X_{+}^{k}
\hspace{5mm} , \hspace{5mm}
[X_{+}^{i},X_{+}^{j}] = 0 \ ,
\eeq
so the gauge-gauge brackets are identified with $\,\mathfrak{g}_{\textrm{gauge}}=\mathfrak{iso(3)}$. As a result, the algebra $\,\mathfrak{g_{0}}\,$ turns out to be
\beq
\label{g_IIA_algebra}
\mathfrak{g_{0}} \,\,=\,\,  \mathfrak{iso(3)} \oplus \mathfrak{u(1)^{6}} \,\,\sim\,\, \mathfrak{so(3)} \oplus \mathfrak{nil_{9}(2)} \ ,
\eeq
where $\,\mathfrak{nil_{9}(2)}\,$ denotes a nilpotent $9$-dimensional ideal of order two (three steps) spanned by the generators $\,\left\lbrace X_{+}^{i} \,,\, Z_{+a} \,,\, Z_{+i} \right\rbrace\,$ and with lower central series
\beq
\left\lbrace X_{+}^{i} \,,\, Z_{+a} \,,\, Z_{+i} \right\rbrace \supset \left\lbrace Z_{+a} \,,\, Z_{+i} \right\rbrace \supset 0 \ .
\eeq
The main property to be highlighted is that there is an unique gauge group, i.e., 
\beq
G_{0} = \textrm{ISO}(3) \ltimes \textrm{U}(1)^{6} \ ,
\eeq
underlying all the solutions of the IIA geometric theory. This was already noted for the supersymmetric solution in ref.~\cite{Dall'Agata:2009gv}. As a final remark, none of the generators in the adjoint representation vanishes at these solutions, so the algebra $\,\mathfrak{g_{0}}\,$ in (\ref{g_IIA_algebra}) is actually embeddable within the $\,\mathfrak{so(6,6)}\,$ duality group. 

The above gauge group has three compact and nine non-compact generators. The latter are spontaneously broken at all critical points. The corresponding vector bosons in such cases acquire a mass due to gauge symmetry breaking by absorbing a scalar degree of freedom. In the scalar mass spectra listed above, there will always be nine scalar fields that  do not correspond to propagating degrees of freedom. Being pure gauge, these do not appear in the scalar potential and hence have $m^2 = 0$.

In all critical points considered above, the number of scalar fields with $m^2 = 0$ exceeds nine. This implies that there will always be a number of propagating degrees of freedom whose value is not fixed by the quadratic terms in $V$. Of course there could be higher-order terms that do give rise to moduli stabilisation, or could lead to a negative potential energy. However, in contrast to the Minkowski case, such scalar fields do not represent a potential instability due to the additional contribution from the space-time curvature. Instead, in Anti-de Sitter one should be worried about fields whose quadratic mass term is at the B.F. bound, and if possible verify if their higher-order terms give rise to stability or rather to tachyons. Having no such mass values in our spectra, this issue plays no role here.

\subsubsection*{The critical point solution of $\dim=2$}

Besides the previous sixteen critical points, the landscape of the $\,\cN=4\,$ type IIA geometric theory still has a $\,\dim=2\,$ piece. In terms of the flux background, it is given by
\beq 
\label{IIA_GKP}
c_{0} = c_{1} = \tilde{c}_{1} = 0
\hspace{5mm},\hspace{5mm}
a_0 = a_1 = 0
\hspace{5mm},\hspace{5mm} 
b_1 = a_2
\hspace{5mm},\hspace{5mm} 
b_0 =-a_3 \ .
\eeq
After three T-dualities along the $\,\eta^{a}\,$ directions, where $\,a=1,3,5$, this type IIA background is mapped to a type IIB one only involving certain gauge fluxes (see table~\ref{table:unprimed_fluxes}). We postpone the discussion of this solution to the next section where type IIB backgrounds including gauge fluxes, O$3$-planes and D$3$-branes will be explored in full generality.

\section{Non-geometric type IIB flux compactifications}
\label{sec:nongeom_IIB}

In this final part we study another realisations of the SO$(3)$-truncation of half-maximal supergravity in four dimensions. This time it will be in the context of isotropic type IIB compactifications on $\mathbb{T}^{6}/(\mathbb{Z}_{2} \times \mathbb{Z}_{2})$ including generalised background fluxes.

\subsection{GKP flux compactifications: stability and gaugings}
\label{sec:extrema_N=4_IIB}

Let us start with the well known type IIB string compactifications including a background for the gauge fluxes $\,(H_{3},F_{3})\,$ and eventually O$3$-planes and/or D$3$-branes sources in order to cancel a flux-induced tadpole 
\beq
\label{Tad3}
\int_{10\textrm{d}} ( H_{3} \wedge F_{3} ) \wedge C_{4}
\hspace{5mm} \Rightarrow \hspace{5mm}
N_{3} = H_{3} \wedge F_{3} \ ,
\eeq
for the R-R gauge potential $\,C_{4}$. These compactifications were presented in the seminal GKP paper of ref.~\cite{Giddings:2001yu} and deeply explored from the moduli stabilisation point of view in refs~\cite{Kachru:2002he,Frey:2002hf,DeWolfe:2004ns,DeWolfe:2005uu} among many others. 

When compatible with an SO($3$) truncation of half-maximal supergravity, these compactifications correspond to having non-vanishing
$\,(a_{0}, \,a_{1}, \,a_{2}, \,a_{3})\,$ as well as $\,(b_{0},\,b_{1}, \,b_{2}, \,b_{3})\,$ flux components in table~\ref{table:unprimed_fluxes}. The flux-induced superpotential for the resulting $STU$-models then reads
\beq
\label{W_IIB_GKP}
W_{\textrm{GKP}}= a_0 - 3 \, a_1 \, U + 3 \, a_2 \, U^2 - a_3 \, U^3 - \left( \,b_0 - 3 \, b_1 \, U + 3 \, b_2 \, U^2 - b_3 \, U^3 \,\right) S \ ,
\eeq 
and the theory comes out with a no-scale structure \cite{Cremmer:1983bf}. It is worth noticing at this point that in these IIB models with only gauge fluxes there are no quadratic constraints from (\ref{QCL}) to fulfill.

At the origin of the moduli space, the potential energy arranges into a sum of square terms hence being non-negative defined
\beq 
\label{Mink_cond}
V_{0}= \frac{1}{32} \left( \,\,(a_0 - b_3)^2 + 3 \,(a_1+b_2)^2 + 3 \, (a_2 - b_1)^2 + (a_3 + b_0)^2 \,\,\right) \ .
\eeq
Using the stabilisation of the imaginary part of the modulus $T$, it can be shown that there is no solution to the extremum conditions without satisfying $\,V_{0}=0\,$, i.e., any solution will be a Minkowski extremum. Then the $\,H_{3}\,$ flux background is related to the $\,F_{3}\,$ one via
\beq 
\label{FH_GKP}
b_3 = a_0 \hspace{5mm},\hspace{5mm} b_2 =- a_1 \hspace{5mm},\hspace{5mm} b_1 = a_2 \hspace{5mm},\hspace{5mm} b_0 =-a_3 \ , 
\eeq
and the flux-induced tadpole in (\ref{Tad3}) simply reads
\beq 
N_{3}= a_{0}^{2} + 3 \, a_{1}^{2} + 3 \, a_{2}^{2} + a_{3}^{2} \ .
\eeq
The $\,\kappa_{1}\,$ and $\,\kappa_{2}\,$ values entering the gravitini mass matrix $\,A^{ij}_{1}\,$ in (\ref{kappa-eigenvalues}), and then determining the amount of supersymmetry preserved at an extremum, are given by
\beq \kappa_{1} = \frac{3}{4 \, \sqrt{2}} \, \sqrt{\left( a_{0}- 3\,
a_{2}\right)^{2} + \left( a_{3}- 3\, a_{1}\right)^{2}}
\hspace{5mm},\hspace{5mm} \kappa_{2} = \frac{3}{4 \, \sqrt{2}} \,
\sqrt{\left( a_{0} + a_{2}\right)^{2} + \left( a_{1} +
a_{3}\right)^{2}} \ . \eeq
As a consequence, a generic GKP solution will be non-supersymmetric. However, let us comment about two interesting limits which give rise to solutions that preserve certain amount of supersymmetry:
\begin{itemize}

\item The first limit is that of taking $\,a_{0} = 3 \, a_{2}\,$ and $\,a_{3} = 3 \, a_{1}$. This limit results in $\,\kappa_{1} = 0\,$ and $\,\kappa_{2}=\frac{3 \sqrt{a_{1}^2 + a_{2}^2}}{\sqrt{2}}\,$ so that the solutions preserve $\mathcal{N}=1$ supersymmetry.

\item The second limit is that of taking $\,a_{0} = - a_{2}\,$ and $\,a_{3} = -a_{1}$. This limit results in $\,\kappa_{2} = 0\,$ and $\,\kappa_{1}=\frac{3 \sqrt{a_{1}^2 + a_{2}^2}}{\sqrt{2}}\,$ so that the solutions preserve $\,\mathcal{N}=3\,$ supersymmetry \cite{Frey:2002hf}.
\end{itemize}

Let us now present the mass spectrum of these $\,\mathcal{N}=4\,$ compactifications\footnote{The numerical values of the eigenvalues of the mass matrix were computed in ref.~\cite{Saltman:2004sn} for some de Sitter GKP examples corresponding to non-isotropic moduli VEVs.}. In terms of the quantities
\beq
\begin{array}{cclc}
M&=&\dfrac{1}{16} \, \Big(\, 9 \, \left( a_1^2 + a_2^2\right) + 6 \, (a_0 \, a_2 + a_1 \, a_3) + 5 \, (a_0^2 + a_3^2) \, \Big)  & , \\[5mm]
N&=&\dfrac{1}{16} \, \Big( \, 5 \, \left( a_1^2 + a_2^2\right) - 2 \,
(a_0 \, a_2 + a_1 \, a_3) + (a_0^2 + a_3^2) \, \Big) & ,
\\[5mm]
Q&=& \dfrac{1}{16} \, \sqrt{ \Big( \,(a_0 - 3 \, a_2)^2 + ( a_3 - 3 \,
a_1)^2 \, \Big) \, \Big( \, (a_0 + a_2)^2 + ( a_1 + a_3)^2 \,
\Big)} & ,
\end{array}
\eeq
the moduli (masses)$^{2}$ as well as their multiplicities are given by
\beq
\begin{array}{c}
M \pm 3 \, Q \,\,(\times 1) \hspace{5mm} , \hspace{5mm} N \pm
Q\,\,(\times 6) \hspace{5mm} , \hspace{5mm} \dfrac{1}{8} \, \left(
\, (a_0 + a_2)^{2} + (a_1 + a_3)^{2} \, \right) \,\,(\times 3)
\hspace{4mm} , \hspace{4mm} 0\,\,(\times 21) \ . \nonumber
\end{array}
\eeq
Only the third of the above masses is not recovered when considering only the scalars of the SO$(3)$ truncation. Clearly though, these solutions can never be stable because of the general presence of flat directions.

The last question we will address is to determine the gauging underlying this GKP backgrounds. The brackets in (\ref{SW_algebra_ZX}) get now simplified to
\beq
\label{brackets_GKP}
\begin{array}{cccc}
\left[ {X_{+}}^{m} , {X_{+}}^{n} \right] =  {\tilde{F}}^{mnp} \,\,Z_{+ p}
&\hspace{5mm},\hspace{5mm}&
\left[  {X_{+}}^{m} , {X_{-}}^{n} \right] =  {\tilde{F}}^{mnp} \,\,Z_{- p} & , \\[2mm]
\left[  {X_{-}}^{m} , {X_{-}}^{n} \right] =  {\tilde{H}}^{mnp} \,\,Z_{- p}
&\hspace{5mm},\hspace{5mm}&
\left[  {X_{-}}^{m} , {X_{+}}^{n} \right] =  {\tilde{H}}^{mnp} \,\,Z_{+ p} & .
\end{array}
\eeq 
Even when there are no quadratic constraints for the fluxes to obey, the antisymmetry of the brackets in (\ref{brackets_GKP}) when substituting (\ref{FH_GKP}) is guaranteed iff
\beq 
\label{antisym_cond} 
Z_{+ a} = - Z_{- i} 
\hspace{5mm} , \hspace{5mm} 
(a_{0} + a_{2}) \, Z_{+ i} = (a_{1} + a_{3}) \, Z_{- i}
\hspace{5mm} , \hspace{5mm} 
(a_{0} + a_{2}) \, Z_{- a} = (a_{1} +
a_{3}) \, Z_{- i} \ , 
\eeq
again with pairs $\,(a,i)=\left\lbrace (1,2),(3,4),(5,6) \right\rbrace$. As a result, the isometry $\,Z_{\alpha m}\,$ generators span a central extension of a $\,\mathfrak{u(1)^{12}}\,$ algebra specified by the $\,X_{\alpha}^{m}\,$ generators in (\ref{brackets_GKP}). Consequently, $\,\textrm{R}_\textrm{Adj}\left[Z_{\alpha m}\right]=0\,$ and the antisymmetry conditions in (\ref{antisym_cond}) are trivially satisfied in this representation\footnote{In other words, the adjoint representation is no longer faithful.}. This is the representation of the gauging which has to be embeddable into the $\mathfrak{so(6,6)}$ duality algebra, so the gauging is the abelian group $\,G_{0}=U(1)^{12}$.

\subsection{Non-geometric backgrounds: the $\textrm{SO}(3,3) \times \textrm{SO}(3,3)$ splitting}

In this final section we move to study some gaugings which cannot be realised as geometric type II string compactifications. Specifically, we will focus on those based on the direct product splitting $\,\textrm{SO}(3,3) \times \textrm{SO}(3,3)\,$ discussed in refs~\cite{deRoo:2002jf, deRoo:2003rm,deRoo:2006ms} and further interpreted as non-geometric flux compactifications in refs~\cite{Roest:2009dq, Dibitetto:2010rg}.

This splitting implies the factorisation of the gauge group in terms of $G_1 \times G_2$, where furthermore $\,G_1\,$ and $\,G_2\,$ were chosen in ref.~\cite{deRoo:2003rm} to be electric and magnetic respectively. This provides the simplest solution to the the second set of quadratic constraints in \eqref{QC} and moreover a non-trivial gauging at angles which is necessary in order to guarantee moduli stabilisation \cite{deRoo:1985jh}. In ref.~\cite{deRoo:2003rm} some de Sitter solutions have been found by investigating the case in which $G_1$ and $G_2$ are chosen to be some SO($p,q$), with $\,p+q=4$. Later on non-semisimple gaugings of the form CSO($p,q,r$)$\,\times\,$CSO($p,q,r$) have been investigated in ref.~\cite{deRoo:2006ms}, but no de Sitter solutions were found.

Let us go deeper into the vacua structure of these CSO($p,q,r$)$\,\times\,$CSO($p,q,r$) gaugings. In order to do so, we will use the parameterisation of the embedding of each CSO factor inside {SO}($3,3$) in terms of the two real symmetric matrices $M_{\pm}$ and $\tilde{M}_{\pm}\,$ as explained in ref.~\cite{Roest:2009tt}. In the case of the SO($3$) truncation, these are given by
\beq 
M_{+} \equiv \textrm{diag}\left(-a'_{0} \,\,,\,\, \tilde{c}_{1}\,,\,\tilde{c}_{1}\,,\,\tilde{c}_{1}\right)
\hspace{5mm},\hspace{5mm} 
\tilde{M}_{+} \equiv \textrm{diag}\left(-a_{0} \,\,,\,\, \tilde{c}'_{1}\,,\,\tilde{c}'_{1}\,,\,\tilde{c}'_{1}\right) \ , 
\eeq
together with
\beq 
M_{-} \equiv \textrm{diag}\left(b'_{3} \,\,,\,\, \tilde{d}_{2}\,,\,\tilde{d}_{2}\,,\,\tilde{d}_{2}\right)
\hspace{5mm},\hspace{5mm}
\tilde{M}_{-} \equiv \textrm{diag}\left(b_{3} \,\,,\,\,\tilde{d}'_{2}\,,\,\tilde{d}'_{2}\,,\,\tilde{d}'_{2}\right) \,\ , 
\eeq
where the relation between the entries of the above matrices and the embedding tensor components can be read off from tables~\ref{table:unprimed_fluxes} and \ref{table:primed_fluxes}. The flux-induced superpotential in (\ref{W_fluxes}) then reduces to
\beq
\label{W_SO(3,3)xSO(3,3)}
\begin{array}{cclcc}
W_{\textrm{SO}(3,3)^{2}} &=& a_0 + b_3 \, S \, U^3 - 3 \, \tilde{c}_{1} \, T \,U  - 3 \, \tilde{d}_{2} \, S  \,T \,U^2 &  & \\[3mm]        
&+& a_0' \,T^3 \,U^3 + b_3' \, S \, T^3  - 3 \, \tilde{c}_{1}' \,T^2 \, U^2   + 3 \, \tilde{d}'_{2} \, S  \,T^2 \,U  &  .
\end{array}
\eeq 

The antisymmetry of the brackets in (\ref{SW_algebra_ZX}) now translates into
\beq
Z_{+i} = X_{+}^{i} = Z_{-a} = X_{-}^{a} = 0 \ , 
\eeq
and the resulting twelve dimensional algebra $\,\mathfrak{g_{0}}\,$ is written as
\beq
\begin{array}{cccl}
\label{SO(3,3)xSO(3,3)_brackets}
[Z_{+a},Z_{+b}] = \phantom{-}\tilde{c}'_{1} \, Z_{+c} - a'_{0} \, X_{+}^{c}    &     \hspace{5mm} , \hspace{5mm}      &  
[Z_{-i},Z_{-j}] = \tilde{d}'_{2} \, Z_{-k} + b'_{3} \, X_{-}^{k} & , \\[2mm]
[Z_{+a},X_{+}^{b}] = \phantom{-}\tilde{c}_{1} \, Z_{+c} + \tilde{c}'_{1} \, X_{+}^{c} &     \hspace{5mm} , \hspace{5mm}      &  
\,\,[Z_{-i},X_{-}^{j}] =  \tilde{d}_{2} \, Z_{-k} + \tilde{d}'_{2} \, X_{-}^{k} & , \\[2mm]
\,\,[X_{+}^{a},X_{+}^{b}] = -a_{0} \, Z_{+c} + \tilde{c}_{1} \, X_{+}^{c}                                 &     \hspace{5mm} , \hspace{5mm}      &  
\,\,\,[X_{-}^{i},X_{-}^{j}] =  b_{3} \, Z_{-k} + \tilde{d}_{2} \, X_{-}^{k} & .
\end{array}
\eeq
The first set of quadratic constraints in (\ref{QC}) gets also simplified and forces the products $\,M_{+} \, \tilde{M}_{+}\,$ and $\,M_{-} \, \tilde{M}_{-}\,$ to be proportional to the identity matrix.

For the sake of simplicity we will consider the case of having only unprimed fluxes, i.e. having a type IIB background including gauge $\,(F_{3},H_{3})\,$ and non-geometric $\,(Q,P)\,$ fluxes. Such backgrounds, although being non-geometric, still admit a locally geometric description and in accord with ref.~\cite{Dibitetto:2010rg}, they can never give rise to semisimple gaugings. Their associated flux-induced superpotential takes the quite simple form of
\beq
\label{W_SO(3,3)xSO(3,3)-unprim}
\begin{array}{ccc}
W_{\textrm{SO}(3,3)^{2}}^{\textrm{loc. geom.}} &=& a_0 + b_3 \, S \, U^3 - 3 \, \tilde{c}_{1} \,T \, U   - 3 \, \tilde{d}_{2} \, S  \,T \,U^2 \ .        
\end{array}
\eeq 
These backgrounds already satisfy all of the quadratic constraints as well as the extremality conditions for the axions at the origin of moduli space\footnote{This fact points out that the origin of moduli space is an especially interesting point even though it is not the most general solution since this flux background is not
duality invariant.}. In addition, their corresponding flux-induced tadpoles are given by
\beq
\label{Tad37} 
N_{3}=a_{0} \, b_{3} \hspace{10mm} , \hspace{10mm} N_{7}=\tilde{N}_{7}=N'_{7}=0 \ , 
\eeq
where $N_{7}$, $\tilde{N}_{7}$ and $N'_{7}$ relate to the SL$(3)$-triplet of $7$-branes in a type IIB S-duality invariant realisation of the theory \cite{Bergshoeff:2006ic,Bergshoeff:2006jj}. In fact, the second condition in (\ref{Tad37}) is actually identified with $\,\cN=4\,$ quadratic constraints since these $7$-branes would break from half-maximal to minimal supersymmetry.

\begin{table}[h!]
\renewcommand{\arraystretch}{1.80}
\begin{center}
\scalebox{0.87}[0.87]{
\begin{tabular}{ | c || c | c | c | c | c | c |}
\hline
\textrm{ID} & $a_{0}$ & $\tilde{c}_{1}$ & $b_{3}$ & $\tilde{d}_{2}$ & $V_0$  & B.F.  \\[1mm]
\hline \hline
$1$ & $-\lambda$ & $\lambda$ & $-\lambda$ & $-\lambda$ & $-\dfrac{3 \lambda^{2}}{8}$ & $m^2=-\dfrac{2}{3}\rightarrow$ stable \\[1mm]
\hline
$2$ & $\lambda$ & $-\lambda$ & $-\lambda$ & $-\lambda$ & $\dfrac{\lambda^{2}}{8}$ & unstable de Sitter\\[1mm]
\hline
$3_a$ & $5 \, \lambda$ & $3 \, \lambda$ & $-\lambda$ & $-\lambda$ & $-\dfrac{15 \lambda^{2}}{8}$ & $m^2=-\dfrac{26}{15}\rightarrow$ unstable \\[1mm]
\hline
$3_b$ & $-\lambda$ & $\lambda$ & $5 \, \lambda$ & $-3 \, \lambda$ & $-\dfrac{15 \lambda^{2}}{8}$ & $m^2=-\dfrac{26}{15}\rightarrow$ unstable \\[1mm]
\hline
\end{tabular}
}
\end{center}
\caption{Set of extrema of the scalar potential (at the origin of the moduli space) for the $\,\textrm{SO}(3,3) \times \textrm{SO}(3,3)\,$ embeddable type IIB backgrounds admitting a locally geometric description. We also present their stability according to the B.F. bound in (\ref{BF_bound}).} 
\label{table:elecxmag}
\end{table}

Restricting our search of extrema to the origin of the moduli space, we find five critical points some of them with novel features compared to the ``geometric'' results obtained in the previous sections. Apart from the GKP-like solution appearing when switching off the non-geometric fluxes, i.e, $\,\tilde{c}_{1}=\tilde{d}_{2}=0\,$, the set of extrema of the scalar potential and their vacuum energy are summarised in table~\ref{table:elecxmag}. Notice that solutions $3_a$ and $3_b$ are related to each other by a simultaneous inversion of the $S$ and $U$ moduli fields, i.e., by an element of the compact subgroup $\,\textrm{SO}(2)^{3}\,$ of the duality group. The critical points labelled by $1$ and $2$ are invariant under this transformation. This is similar to the $\mathbb{Z}_2 \times \mathbb{Z}_2$ structure in the geometric IIA case. However, in contrast to that situation, the other critical points in table~\ref{table:elecxmag} cannot be related by non-compact duality transformations. Therefore these are solutions to different theories.

The computation of the gravitini mass matrix $\,A^{ij}_{1}\,$ in (\ref{kappa-eigenvalues}) shows that the solution $1$ in table~\ref{table:elecxmag} preserves $\,\cN=4\,$ supersymmetry whereas all the others turn out to be non-supersymmetric. The normalised mass spectra for these solutions are as follows:
\begin{itemize}

\item The normalised masses and their multiplicities for the solution $1$ are given by
\beq
\dfrac{4}{3} \,\,(\times \, 2) \hspace{5mm} , \hspace{5mm} 0 \,\,(\times \, 24) \hspace{5mm} , \hspace{5mm} -\dfrac{2}{3} \,\,(\times \, 12) \ . 
\eeq
The twelve tachyonic modes imply $\,m^2=-2/3\,$ and then satisfy the B.F. bound in (\ref{BF_bound}) ensuring the stability of this AdS$_4$ solution.

\item The normalised masses and their multiplicities for the solution $2$ are given by
\beq
6 \,\,(\times \, 10) \hspace{5mm} , \hspace{5mm} 4 \,\,(\times \, 18) \hspace{5mm} , \hspace{5mm} -2 \,\,(\times \, 2) \hspace{5mm} , \hspace{5mm} 0 \,\,(\times \, 8) \ , 
\eeq
so this de Sitter solution is automatically unstable since it contains two tachyons.

\item The normalised masses and their multiplicities for the solutions $3_{a,b}$ are given by
\beq
\begin{array}{cccccccc}
-\dfrac{26}{15} \,\,(\times 5)  & \hspace{3mm},\hspace{3mm} & -\dfrac{4}{5}\,\,(\times 9) & \hspace{3mm},\hspace{3mm} & -\dfrac{2}{15}\,\,(\times 1) & \hspace{3mm},\hspace{3mm} &  \dfrac{1}{15} \left( \, 23 \pm \sqrt{1009} \, \right)\,\,(\times 1) & , \\[4mm]
\dfrac{2}{5}\,\,(\times 5) & , &   \dfrac{16}{15}\,\,(\times 1)  & , &  \dfrac{4}{3}\,\,(\times 9)   & , &  0\,\,(\times 6) & ,
\end{array}
\nonumber 
\eeq
so these AdS$_{4}$ solutions do not satisfy the B.F. bound in (\ref{BF_bound}) for fourteen tachyonic modes hence becoming unstable.

\end{itemize}
\noindent
We would like to point out that in these non-geometric flux vacua the lightest mode generically no longer belongs to the SO($3$) truncation. 

Concerning the gauge group underlying these locally geometric type IIB backgrounds, it is directly identified with
\beq
\textrm{G}_{0} = \textrm{ISO}(3) \times \textrm{ISO}(3) \ ,
\eeq
when keeping only unprimed fluxes in the brackets of (\ref{SO(3,3)xSO(3,3)_brackets}). The three different theories correspond to inequivalent embeddings of this gauge group in the global symmetry group. All critical points break the non-compact generators of this gauge group, and hence six of the massless scalars in the mass spectra listed above correspond to non-physical scalars.

As a final remark, we want to highlight that table \ref{table:elecxmag}, even though not being exhaustive, contains interesting solutions such as an example of $\mathcal{N}=4$ supersymmetric Anti-de Sitter vacuum and an example of de Sitter solution obtained from a non-semisimple gauging. The latter is the first example with such a gauge group; all previously constructed de Sitter solutions are based on semi-simple groups \cite{deRoo:2002jf, deRoo:2003rm}.

\section{Conclusions}

We have presented a general method for an exhaustive analysis of the
vacua structure of isotropic $\mathbb{Z}_2 \times \mathbb{Z}_2$ flux
compactifications, and applied it to various cases with a single
set of sources. These vacua correspond to critical points of the
SO($3$) truncation of \mbox{$\mathcal{N}=4$} gauged supergravity. Moreover,
we have presented the explicit dictionary needed to relate such
half-maximal supergravity theories to $\cN = 1$ theories constructed
by a given superpotential. Finally, in appendix~\ref{App:N1_vacua},
we present the general vacuum structure of the type IIA geometric
theory in the presence of sources compatible with $\cN = 1$
supersymmetry.

One of the main results of this paper is the proof that all
geometric IIA vacua belong to a single theory with gauge group
$G_{0} = \textrm{ISO}(3) \ltimes \textrm{U}(1)^{6}$. Of the four
AdS$_4$ critical points of this theory, one is supersymmetric. The
other three are non-supersymmetric and nevertheless two of them are
perturbatively stable. The above statement is actually true up to
the $\mathbb{Z}_2 \times \mathbb{Z}_2$ symmetry presented in
section~\ref{sec:geom_IIA}. Furthermore, our full analysis of these
geometric IIA compactifications leads us to conclude that no de
Sitter solutions are present in the $\mathcal{N}=4$ theory, whereas they
are present for $\cN = 1$. These were already found in refs~\cite{Caviezel:2008tf, deCarlos:2009qm}, and we show in the appendix~\ref{App:N1_vacua} that they are in fact the only de Sitter for such compactifications.

For type IIB compactifications, the full set of vacua has been studied in the presence of only gauge fluxes. 
We provided some relevant examples of solutions to the half-maximal
theory describing a non-geometric type IIB background. The gauge
group in this case is always $\textrm{ISO}(3) \times
\textrm{ISO}(3)$; however, the different critical points belong to
inequivalent embeddings of the gauge group within SO($6,6$) and
hence different theories. Amongst the critical points of these theories we found a new
unstable de Sitter solution.

It would be interesting to better understand some of the surprising
features of the geometric IIA compactification that follow from our
classification. Why does this lead to a unique theory with moduli
stabilisation, at least in the SO$(3)$ truncation? Similarly, it is
intriguing that this truncation captures the scalars that are relevant
for the stability analysis; all other scalars have positive masses. Can
one understand why this happens in the present case, and not in
e.g.~SO$(8)$ gauged maximal supergravity? Another difference with that
theory is that the supersymmetric vacuum is not the one with highest
energy. These are amongst the open questions that deserve further study.
Finally, if possible it would be very interesting to perform a similar
classification for the general non-geometric IIB compactifications. The
few examples that we presented in this paper already indicate that the
landscape of these vacua is much richer.

\bigskip

\noindent
{\bf Note added:} Upon completion of this manuscript we received the preprint of ref.~\cite{Aldazabal:2011yz} which has some overlap with parts of the present paper.

%
%

\vspace*{5mm}
\noindent
{\bf \large Acknowledgments}
\vspace*{3mm}

\noindent
We are grateful to A. Borghese, P. C\'amara, A. Rosabal and T. Van Riet for very interesting discussions and R. Linares for collaboration in the early stages of this project. Furthermore, D.R.~would like to express his gratitude to the LMU M\"{u}nchen for its warm hospitality while part of this project was done. The work of the authors is supported by a VIDI grant from the Netherlands Organisation for Scientific Research (NWO).

%
%

%
\appendix

\section{Full $\,\mathcal{N}=1\,$ flux vacua of geometric type IIA}
\label{App:N1_vacua}

The techniques developed to analyse the vacua of the $\cN=4$ theory turn out to be powerful enough to also work out the complete set of solutions of type IIA geometric backgrounds compatible with minimal supersymmetry.
As we saw in section \ref{sec:N=1_form}, the SO($3$) truncation admits an
$\mathcal{N}=1$ superpotential formulation. In this context it
becomes natural to relax the quadratic constraint in (\ref{N6_orth}) which can be
understood as the lack of D$6$-branes orthogonal to the O$6$-planes.
Namely,
\beq 
\label{QC_relax} 
N_{6}^{\bot} = - a_{3}\, c_{0} - a_{2} \, (2\,
c_1 - \tilde{c}_{1}) \neq 0 \ . 
\eeq
After this, the theory no longer enjoys $\mathcal{N}=4$ supersymmetry but it
 still admits an $\mathcal{N}=1$ description\footnote{Nevertheless, any
solution of the $\mathcal{N}=1$ theory compatible with the absence
of such sources can be embedded into the $\mathcal{N}=4$ theory.}.
In this section we will explore its vacuum structure.

We will distinguish between two types of IIA geometric flux
backgrounds, namely, those having only gauge fluxes and those with
both gauge and metric fluxes.

\subsection*{Backgrounds only with gauge fluxes}

Let us start by fixing the components of the metric $\omega$ flux to zero, namely,  
\beq
b_{1}=c_{1}=\tilde{c}_{1}=0 \ .
\eeq
Putting together the first and the second quadratic constraints in (\ref{QC_IIA}) and the extremality conditions, and using again the GTZ algebraic method of prime decomposition (details explained
in section~\ref{sec:alg_geom}), we obtain a solution space consisting of two pieces:

\begin{table}[h!]
\renewcommand{\arraystretch}{1.80}
\begin{center}
\scalebox{0.85}[0.85]{
\begin{tabular}{ | c || c | c |c | c | c | c |c | c | c | c |}
\hline
\textrm{ID} & $a_{0}$ & $a_{1}$ & $a_{2}$ & $a_{3}$ & $b_{0}$ & $b_{1}$ & $c_{0}$ & $c_{1}=\tilde{c}_{1}$ & $V_{0}$ & B.F. \\[1mm]
\hline \hline
$1$ & $0$ & $\dfrac{3 \lambda}{2}$ & $0$ & $\dfrac{5 \lambda}{2}$ & $-\lambda$ & $0$ & $\lambda$ & $0$ & $-\dfrac{3 \lambda^{2}}{32}$ & $m^2=-\dfrac{2}{3}\rightarrow$ stable\\[1mm]
\hline
$2$ & $0$ & $-\dfrac{3 \lambda}{2}$ & $0$ & $\dfrac{5 \lambda}{2}$ & $-\lambda$ & $0$ & $\lambda$ & $0$ & $-\dfrac{3 \lambda^{2}}{32}$ & $m^2=-\dfrac{2}{3}\rightarrow$ stable\\[1mm]
\hline
$3$ & $0$ & $\sqrt{6} \lambda$ & $0$ & $5 \lambda$ & $-4 \lambda$ & $0$ & $\lambda$ & $0$ & $-\dfrac{\lambda^{2}}{4}$ & $\min$\\[1mm]
\hline
$4$ & $0$ & $-\sqrt{6} \lambda$ & $0$ & $5 \lambda$ & $-4 \lambda$ & $0$ & $\lambda$ & $0$ & $-\dfrac{\lambda^{2}}{4}$ & $\min$\\[1mm]
\hline
$5_{s_{1}}$ & $0$ & $s_{1}\,\lambda$ & $\lambda$ & $-2 \,s_{1}\, \lambda$ & $s_{1}\,\lambda$ & $0$ & $-s_{1}\,\lambda$ & $0$ & $-\dfrac{\lambda^{2}}{16}$ & $\min$\\[1mm]
\hline
$6_{s_{1}}$ & $0$ & $s_{1}\,\dfrac{7 \lambda}{3}$ & $-\dfrac{\lambda}{3}$ & $-s_{1}\,\dfrac{14 \lambda}{3}$ & $s_{1}\,\dfrac{11 \lambda}{3}$ & $0$ & $-s_{1}\,\lambda$ & $0$ & $-\dfrac{11 \lambda^{2}}{48}$ & $m^2=-0.14251 \rightarrow$ stable\\[1mm]
\hline
\end{tabular}
}
\end{center}
\caption{The set of stable AdS$_{4}$ extrema of dimension $1$ in the $\mathcal{N}=1$ type IIA theory only with gauge fluxes.} 
\label{table:N=1_vacua_c11=0}
\end{table}

\begin{itemize}

\item[$i)$] The first piece has dimension $2\,$ and it is directly identified with the solution in (\ref{IIA_GKP}) of the $\,\mathcal{N}=4\,$ theory.

\item[$ii)$] The second piece consists of eight critical points of dimension $1$, all of them implying a non-vanishing tadpole for both 
\beq
N_{6}^{\bot} = - a_{3} \, c_{0} \neq 0 \hspace{10mm} \textrm{ and } \hspace{10mm} N_{6}^{||}= - a_{3} \, b_{0} \neq 0 \ , 
\eeq
so they cannot be embedded into the previous $\,\mathcal{N}=4\,$ theory. These moduli solutions are stable AdS$_{4}$ vacua which are summarised in table~\ref{table:N=1_vacua_c11=0}. Finally, these solutions of the $\mathcal{N}=1$ theory are non-supersymmetric except that labelled by $\,1\,$ in table~\ref{table:N=1_vacua_c11=0} which turns out to preserve $\,\mathcal{N}=1\,$ supersymmetry. The scalar potential induced by the fluxes of solutions $2$ and $4$ is respectively  related to that one induced by the fluxes of $1$ and $3$ in table~\ref{table:N=1_vacua_c11=0} by the transformation
\beq
\label{I_parity}
\begin{array}{cccccccc}
\alpha_{3}: & V(S,T,U \,;\, a_{1}, \,\,f_{i}) &=& - i \,\, V(\, i \, S, i \, T,-i \, U \,;\, -a_{1}, \,\,f_{i}) & , 
\end{array}
\eeq
where $f_i$ refers to all the fluxes left invariant. Such a transformation can also be viewed at the level of the superpotential as $\,W(S,T,U) \rightarrow i \,W(S,T,U)$. Unlike those in the previous section, this transformation modifies the K\"ahler potential and, as a consequence, the mass spectrum for the solutions $1$ and $2$ (also $3$ and $4$) is different even when they share the lightest mass. They correspond to completely different solutions although they look quite similar to each other.
\end{itemize}

\subsection*{Backgrounds with both gauge and metric fluxes}

Let us now allow for backgrounds with non-vanishing metric
fluxes. Putting again together the first and second quadratic constraints
in (\ref{QC_IIA}) and the extremum conditions,
and running the GTZ method of prime decomposition, we obtain two
prime factors of dimension $2$ compatible with real fluxes:

\begin{itemize}

\item[$i)$] The first piece represents a branch of non-supersymmetric solutions which cannot be embedded into the $\,\mathcal{N}=4\,$ theory (all the solutions come out with $\,N_{6}^{\bot}\neq 0$). This piece implies $\,a_{0}=a_{1}=0\,$. Without loss of generality, we can set the global scale of $\,V\,$ by fixing $\tilde{c}_{1}=1$ in order to exhaustively explore the structure of extrema by varying the quantity $\,\delta \equiv \left| c_0\right|$. It is found to contain an unstable Minkowski solution \cite{deCarlos:2009qm} at the critical value $\,\delta_{c} \sim 2.69\,$ as well as unstable dS ones if going beyond this critical value (the region with $\,\delta > \delta_{c}\,$ presents an asymptotic behaviour). This is depicted in figure~\ref{fig:plot_dS}.

\begin{figure}[h!]
\centering
\includegraphics[width=7.5cm]{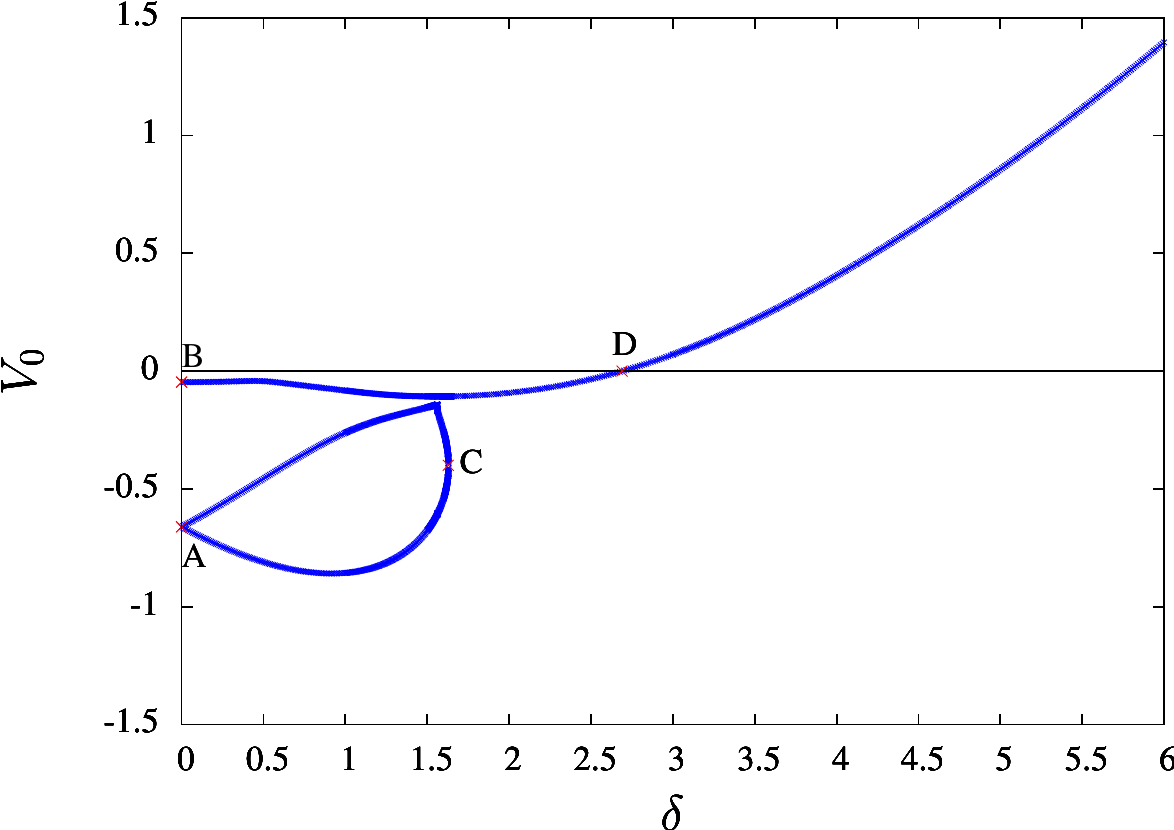}
\includegraphics[width=7.5cm]{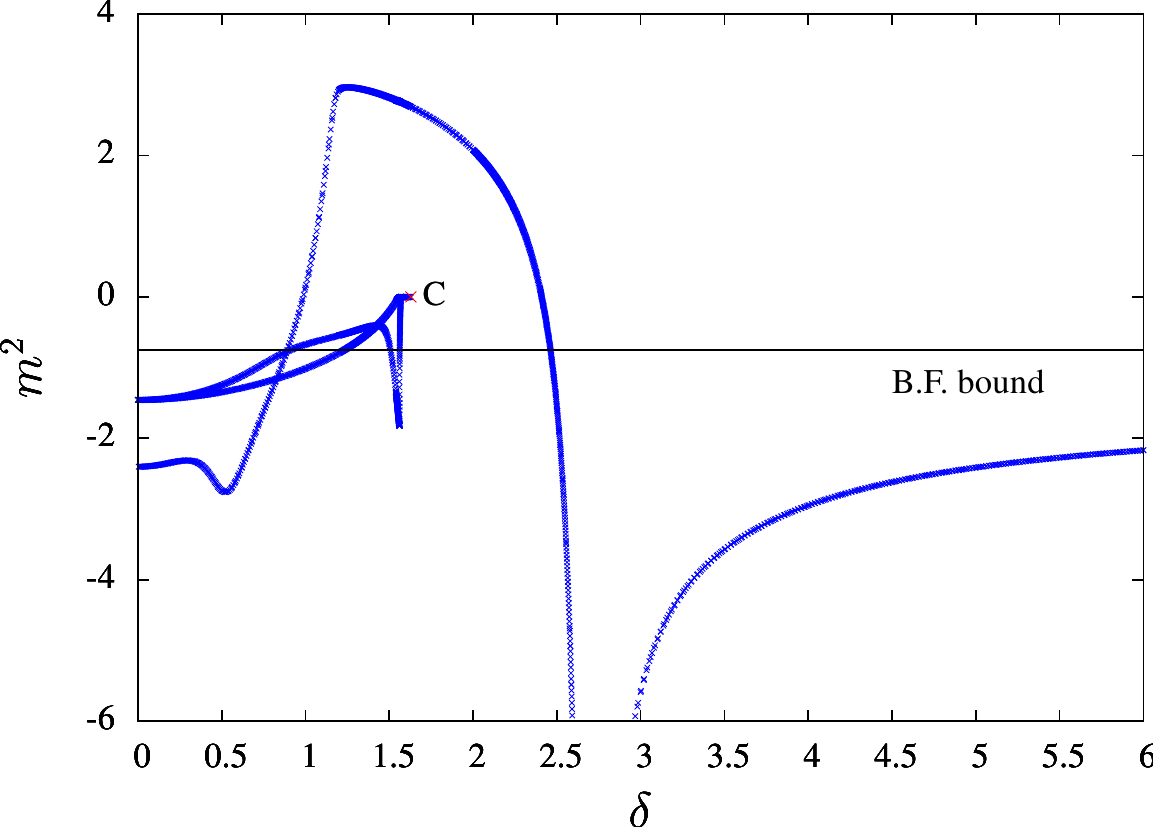}
\caption{Left: Plot of the potential energy at the extrema,
$\,V_{0}\,$, as a function of the scanning parameter $\,\delta$: the
point A corresponds to two degenerate and unstable AdS$_{4}$
solutions; points B and C correspond to singular points; point D
associated to $\delta_{c}\sim 2.69$ is an unstable Minkowski
solution. Right: Plot of the lowest normalised mass in \eqref{mass} as a
function of the scanning parameter $\,\delta$. After reaching the dS
region, the system undergoes an asymptotic behaviour where
$\,m^{2}\rightarrow -\frac{4}{3}\,$ as long as $\,\delta
\rightarrow \infty$.} \label{fig:plot_dS}
\end{figure}

\begin{figure}[h!]
\centering
\includegraphics[width=7.5cm]{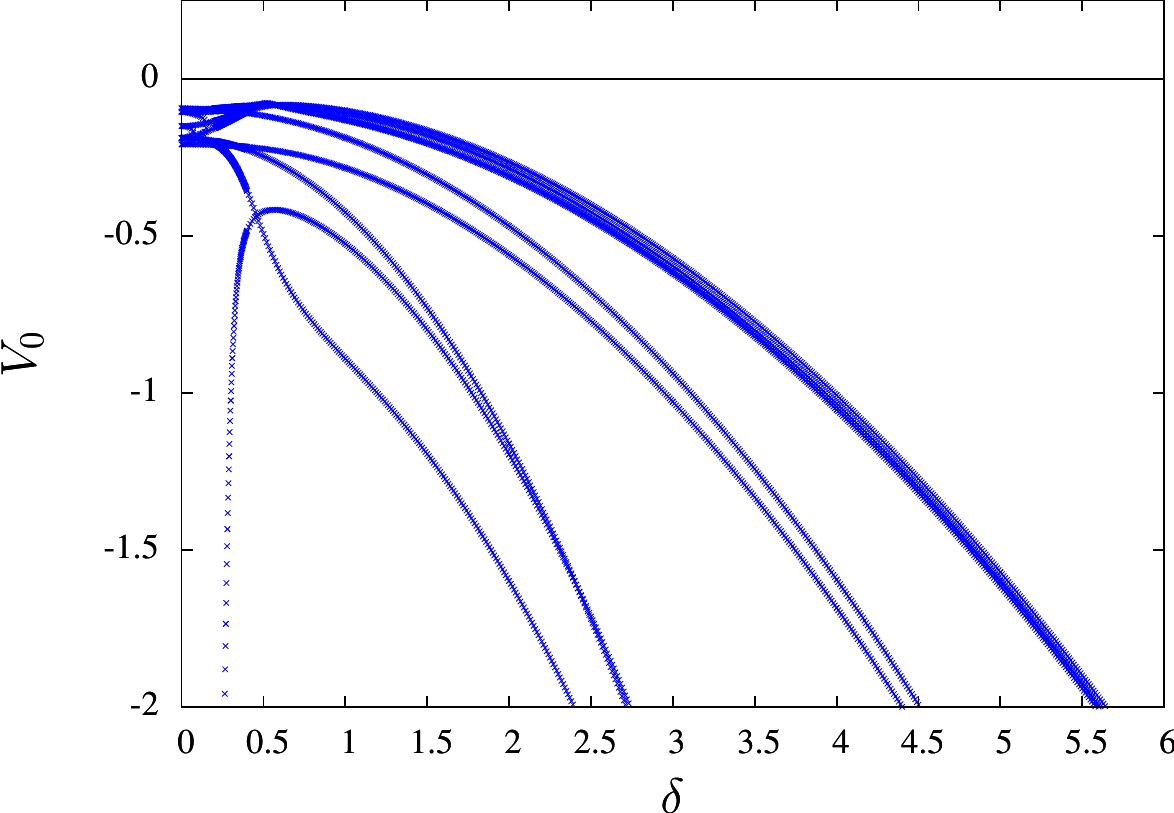}
\includegraphics[width=7.5cm]{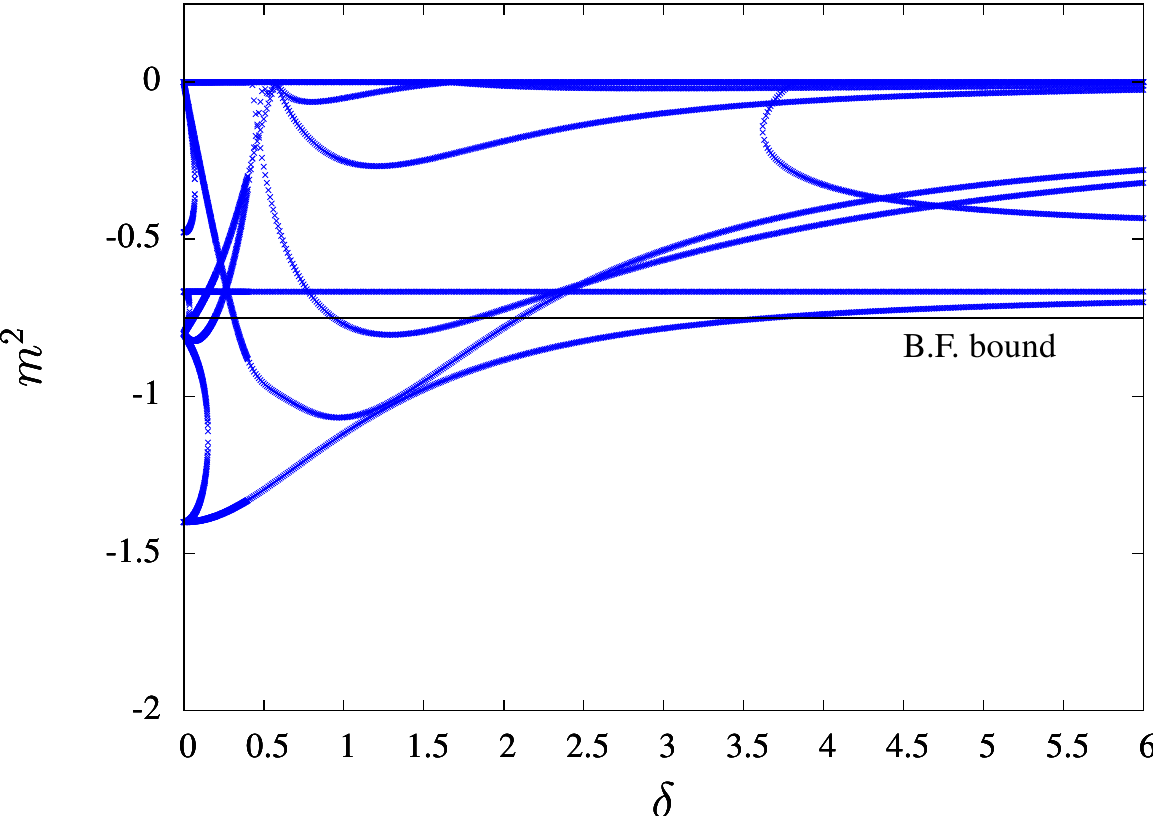}
\caption{Left: Plot of the potential energy at the extrema,
$\,V_{0}\,$, as a function of the scanning parameter $\,\delta$.
Right: Plot of the lowest normalised mass as a function of
the scanning parameter $\,\delta$. As long as
$\,\delta\rightarrow\infty$, the system undergoes a four-fold
asymptotic behaviour with $\,m^{2}\,$ always above the B.F.
bound.} \label{fig:plot_AdS}
\end{figure}

\item[$ii)$] The second piece can also be explored in terms of the quantity $\,\delta \equiv \left| c_0\right|$ after fixing again the global scale of $\,V\,$ by the choice $\,\tilde{c}_{1}=1$. It only contains AdS$_{4}$ solutions which are mostly non-supersymmetric\footnote{The $\,\mathcal{N}=4\,$ quadratic constraints (after relaxing (\ref{QC_relax})) together with the vanishing of the F-terms imply $\,a_{0}=\frac{3}{2}\tilde{c}_{1}$, $\,a_{1}=\frac{3}{2}c_{0}$, $\,a_{2}=-\frac{1}{6}\tilde{c}_{1}$, $\,a_{3}=\frac{5}{2}c_{0}$, $\,b_{0}=-c_{0}$, $\,b_{1}=\frac{1}{3}\tilde{c}_{1}\,$ and $c_{1}=\tilde{c}_{1}$. As a result, for a given value of $\,(c_{0} \,,\, \tilde{c}_{1})\,$, one extremum is always supersymmetric whereas the others (solving $\partial V=0$) are not. At the supersymmetric extremum $\,N_{6}^{\bot}=N_{6}^{||}\,$ holds and $\,m^2=-\frac{2}{3}$. Furthermore, this supersymmetric extremum can be embedded into the $\,\mathcal{N}=4\,$ theory (even $\,\mathcal{N}=8$) when $\,\frac{c_0}{\tilde{c}_{1}}=\frac{1}{\sqrt{15}}\,$ since $\,N_{6}^{\bot}=N_{6}^{||}=0$.} and cannot be embedded into the $\,\mathcal{N}=4\,$ theory because of $\,N_{6}^{\bot}\neq 0$. Nevertheless, some special AdS$_{4}$ solutions with $\,N_{6}^{\bot} = 0\,$ do appear at the special values $\,\delta=0\,$, $\,\delta=1/\sqrt{15}\,$ and $\,\delta=1/\sqrt{3}$, hence being embeddable into the $\,\mathcal{N}=4\,$ theory. This is depicted in figure~\ref{fig:plot_AdS}.

\end{itemize}

%
%

\bibliography{references}
\bibliographystyle{JHEP}

\end{document}